\documentclass{article}
\usepackage{jheppub}
\usepackage{physics}
\usepackage{tikz}

\newcommand\myrenyi[0]{R\'enyi }

\begin{document}

\title{String Stars in Anti de Sitter Space}

\author{Erez Y.~Urbach}
\affiliation{Department of Particle Physics and Astrophysics, Weizmann Institute of Science, Rehovot, Israel}
\emailAdd{erez.urbach@weizmann.ac.il}
\abstract{We study the `string star' saddle, also known as the Horowitz-Polchinski solution, in the middle of $d+1$ dimensional thermal AdS space. We show that there's a regime of temperatures in which the saddle is very similar to the flat space solution found by Horowitz and Polchinski. This saddle is hypothetically connected at lower temperatures to the small AdS black hole saddle. We also study, numerically and analytically, how the solutions are changed due to the AdS geometry for higher temperatures. Specifically, we describe how the solution joins with the thermal gas phase, and find the leading correction to the Hagedorn temperature due to the AdS curvature. Finally, we study the thermodynamic instabilities of the solution and argue for a Gregory-Laflamme-like instability whenever extra dimensions are present at the AdS curvature scale.}
\maketitle

\section{Introduction and Results}\label{sec:intro}
	
	\begin{figure}[t] 
	\centering
		\includegraphics[width=.9\linewidth]{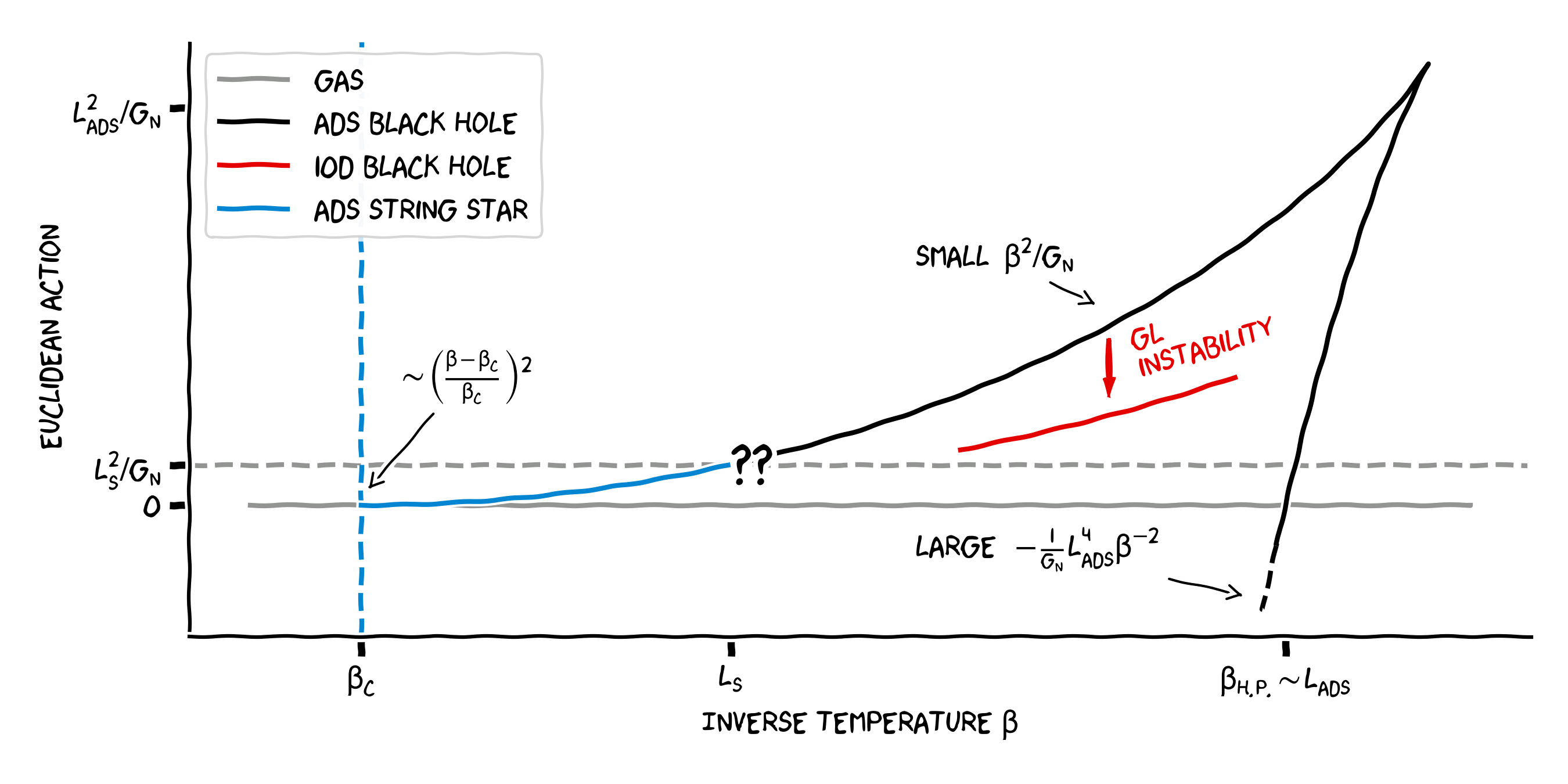}
		\caption{\label{fig:phase_map}
		Schematic phase diagram of the holographic thermal partition function on $S^{d-1}\times S^1$, drawn for $d=3$. 
		$\beta$ labels the asymptotic length of the $S^1$, it is the circumference of the thermal circle at the origin of AdS space in the thermal AdS solution with the same boundary conditions.
		In black is the $d+1$ dimensional AdS black hole.
		For temperatures $l_s \ll \beta \ll l_{ads}$ $10$ dimensional black hole solutions exist (in red). For high enough temperatures small AdS black holes decay into $10$ dimensional black holes through the Gregory-Laflamme (GL) instability (see section \ref{sec:therm_inst}).
		The AdS string star solution (in blue) exists around the Hagedorn temperature. It reliably joins with the thermal gas phase (in gray) at the AdS Hagedorn temperature $\beta_c$. At lower temperatures it is believed to join with the (small) AdS black hole saddle.
		}
	\end{figure}
	One of the important standing goals in quantum gravity is to understand the microstructure of black holes. As a microscopic theory of quantum gravity, different attempts were made to employ string theory to study the microscopic properties of black holes (for reviews see for example \cite{Maldacena:1996ky,Peet:2000hn}).
	One such path is the ``correspondence principle" \cite{Susskind:1993ws,Horowitz:1996nw}. The basic observation is that when one adiabatically shrinks a black hole horizon to the string scale, its thermodynamic properties are qualitatively the same as a generic string state with the same energy \cite{Sen:1995in,Giveon:2006pr}. 
	A canonical version of the correspondence principle was offered in \cite{Horowitz:1997jc} in terms of the thermal string theory partition function on (asymptotically) $R^d\times S^1$ (see also \cite{Damour:1999aw,Khuri:1999ez,Kutasov:2005rr,Giveon:2005jv,Chen:2021emg,Brustein:2021cza,Chen:2021dsw}). For near-Hagedorn temperatures $\beta-\beta_H \ll l_s$, the first string winding mode $\chi$ around the thermal circle is parametrically lighter than the string scale. Upon compactifying the thermal circle, the authors found a ($d$ dimensional) bound state solution of the winding scalar together with gravity. This solution seems to describe a self-gravitating bound state of hot strings. We will call this solution a ``string star''. 
	The Euclidean Schwarzschild black hole is another saddle that contributes to the thermal partition function.
	\footnote{For 
	a recent discussion of winding modes in solutions with this topology see \cite{Brustein:2021qkj}.} The solution is perturbative only for large horizons, when the temperature is low enough, $\beta \gg l_s$. As a result, at generic intermediate temperatures $\beta \sim l_s$ both the black hole and the string star descriptions fail. A naive extrapolation from the perturbative regimes to $\beta \sim l_s$ surprisingly shows a qualitative agreement on the thermodynamic properties between the two saddles. The conclusion might be that thermodynamically, the string star and the black hole are two continuously connected phases. Recently \cite{Chen:2021dsw} gave an extended and modern outlook on the string star solution, including an analysis from the worldsheet perspective. Below we will follow their conventions.

	In this work we study the thermodynamic properties of the string star solution in the middle of thermal anti de Sitter space (AdS), termed the ``AdS string star''. This solution should be understood as a saddle that contributes to the string theory partition function on asymptotically (Euclidean) AdS. This partition function also has a (well known) Euclidean AdS Schwarzschild solution called the AdS black hole. We will argue below that a similar ``correspondence principle" can be qualitatively drawn between the two Euclidean AdS saddles.\footnote{For the Hagedorn temperature in AdS and its relation to string-size AdS black holes see \cite{Barbon:2001di,Barbon:2004dd,Giveon:2005mi,Lin:2007gi,Berkooz:2007fe,Mertens:2015ola,Ashok:2021vww}.}
	Studying string theory on AdS has two main advantages. First, the thermal partition function in asymptotically flat space is an ill defined concept in quantum gravity in general, and also in string theory \cite{Atick:1988si}.
	The AdS partition function on the other hand is well defined, where the radius of the thermal circle is held fixed only at the conformal boundary of space. As a result, the partition function is not strictly thermal in the bulk. The second advantage is that string theory on AdS is holographically dual to a (strongly coupled) conformal field theory (CFT) in one dimension lower \cite{Maldacena:1997re}. Instead of an ill defined thermal partition function, the AdS partition function is well defined and equals the thermal partition function of the dual CFT.

	Let us describe our construction in more details. We consider the string theory Euclidean partition function on asymptotically $EAdS_{d+1}\times X_{9-d}$ with a conformal boundary of $S^{d-1}\times S^1$. Here $X_{9-d}$ is some $9-d$ dimensional compact Euclidean manifold. Holographically we are calculating the thermal partition function on an $S^{d-1}\times S^1$ of a $d$ dimensional CFT. In terms of this holographic CFT, we take the $S^{d-1}$ radius to be $1$, and the length of the $S^1$ to be (dimensionless) $\beta_\text{CFT}$. For the majority of the paper the holographic interpretation won't be crucial. The bulk saddle we will consider is a stringy excitation of $d+1$ dimensional thermal AdS. Thermal AdS has a topology of $R^d \times S^1$. The length of the thermal circle at the origin of the spatial slice is $\beta = l_{ads} \cdot \beta_{CFT}$, with $l_{ads}$ the AdS curvature scale. Far from the origin, the thermal circle grows exponentially with the radial coordinate. But for $\beta-\beta_H \ll l_s$ and close enough to the origin, the winding mode of the string on the thermal circle is light. In this regime we can follow \cite{Horowitz:1997jc} and dimensionally reduce on the time direction. The result is an effective $d$ dimensional theory (on a spatial slice of $AdS_{d+1}$) for the winding mode $\chi$ and gravity. In this setting we look for bound state solutions.\footnote{This is different from the setting of \cite{Brustein:2022uft}, which involves a thermal circle outside of AdS space.} The general properties of this construction are explained in section \ref{sec:properties}.

	\begin{figure}[t] 
	\centering
		\includegraphics[width=.8\linewidth]{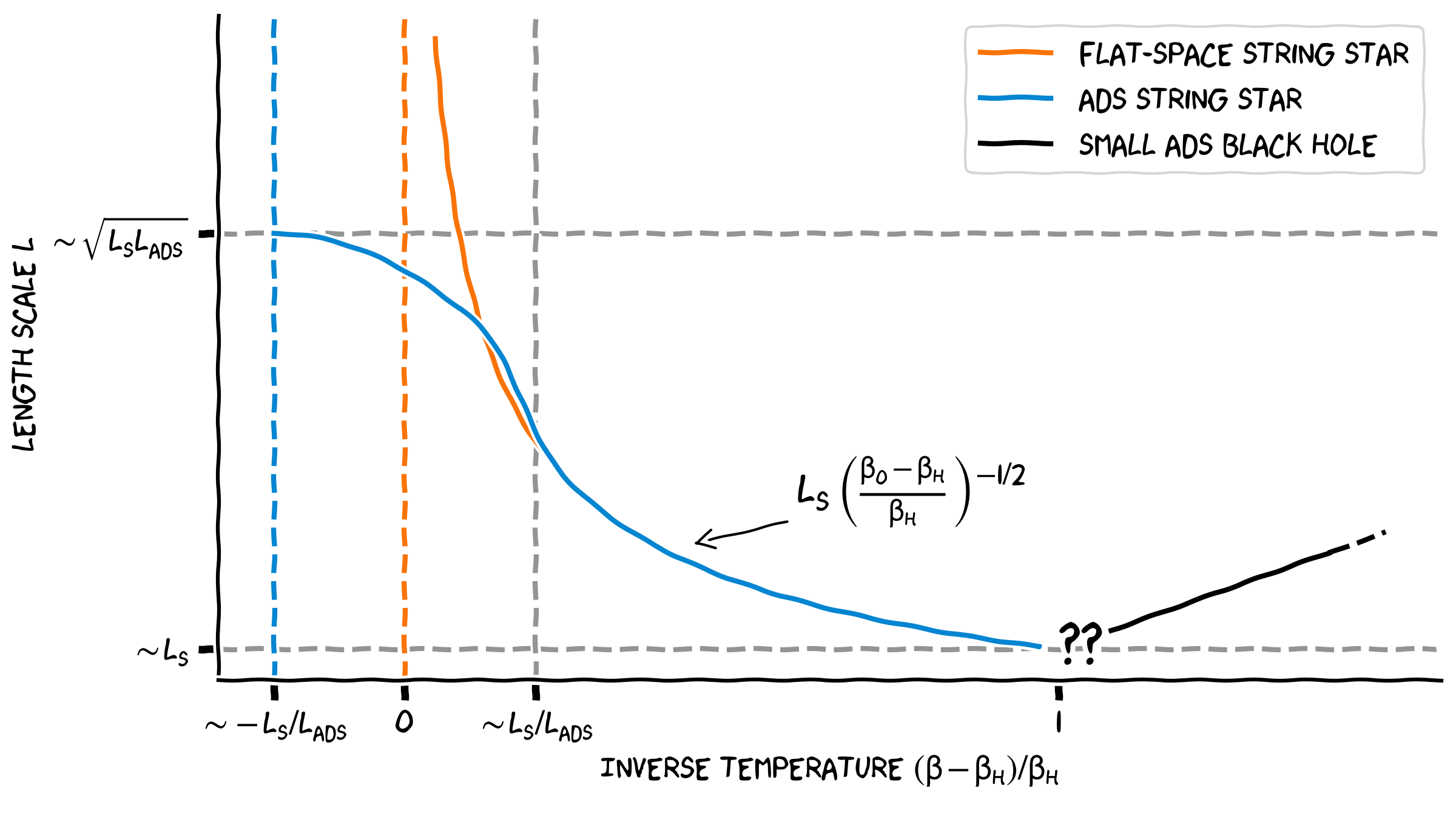}
		\caption{\label{fig:length_scale}
		Schematic picture of the string star solution's length scale as a function of the inverse temperature $(\beta-\beta_H)/\beta_H$ for $3\le d \le 5$. In orange we see the flat space solution of \cite{Horowitz:1997jc}. In blue the AdS solution. The two coincide for low temperatures. They also coincide with the small (AdS) black hole (drawn in black) at the correspondence point. The blue vertical line is the AdS Hagedorn temperature $\beta_c$.
		}
	\end{figure}

	The main results of our analysis are drawn in figures \ref{fig:length_scale} and \ref{fig:phase_map}. In figure \ref{fig:length_scale} we schematically draw the length scale $L$ of the AdS string star compared to the flat space solution of \cite{Horowitz:1997jc} as a function of the temperature $\beta$.
	For low enough temperatures $l_s^2 / l_{ads} \ll \beta-\beta_H \ll l_s$ the size of the two solutions is very small and, by the equivalence principle, the solutions coincide (see section \ref{sec:small_sol}). At higher temperatures the flat space string star grows until its length scale diverges at the Hagedorn temperature $\beta_H$. The AdS string star size also grows with the temperature, but its length scale is bounded by $L\le L_c \sim \sqrt{l_s l_{ads}}$. The solution reaches its maximal size at a critical temperature above the flat space Hagedorn temperature $\beta_c < \beta_H$. At $\beta_c$ the solution's amplitude goes to zero and it joins with the trivial solution $\chi=0$. As a result, $\beta_c$ is also the Hagedorn temperature in AdS space.
	In section \ref{sec:gas_transition} we analytically find the AdS Hagedorn temperature $\beta_c$ at leading order of the AdS curvature. We also find the the solution's profile close to $\beta_c$.

	In figure \ref{fig:phase_map} we schematically plotted the phase diagram of string theory on (asymptotically) thermal $AdS_{d+1}$ as a function of the temperature (see also section \ref{sec:thermo}). The thermal AdS saddle, or the thermal gas phase, (in gray) dominates the canonical ensemble at low temperatures. At high temperatures $\beta \lesssim l_{ads}$ AdS black holes solutions (in black) exist in two branches, ``large'' and ``small''. The large AdS black hole dominates the canonical ensemble at temperatures above the Hawking-Page temperature $\beta_\text{H.P.} \sim l_{ads}$ \cite{Hawking:1982dh,Witten:1998qj,Witten:1998zw}. The small AdS black hole is always metastable. Its size gets smaller with the temperature such that for $l_s \ll \beta \ll l_{ads}$ the solution is approximately a $d+1$ dimensional Schwarzschild black hole. As we said above, an AdS string star solution (in blue) exists for $\beta-\beta_H \ll l_s$, and is approximately the flat space solution for low temperatures in this range.
	Therefore the extrapolation of the AdS string star and the small AdS black hole to $\beta \sim l_s$ follows exactly the flat space analysis of \cite{Horowitz:1997jc}. We find the same qualitative agreement here between the two AdS phases. At higher temperatures the string star's amplitude goes to zero and the solution smoothly joins with the thermal gas phase at $\beta_c$. 
	Note that in this regime of temperatures the gas, the string star and the small AdS black hole are all metastable phases of the canonical ensemble. The small AdS black hole saddle is also known to be thermodynamically unstable \cite{gross:1982}, and in particular suffers from a Gregory-Laflamme (GL) instability due to the full $10$ dimensional geometry \cite{Gregory:1993vy,Hubeny:2002xn,Buchel:2015gxa}. In section \ref{sec:therm_inst} we give evidence that the AdS string star is also thermodynamically unstable, and has a GL-like instability.

	String stars in flat space exist only for $3\le d \le 5$ \cite{Horowitz:1997jc}. In AdS on the other hand string star solutions exist close to $\beta_c$ at every $d$. 
	For $d \ge 6$ the solution turns unreliable as one decreases the temperature to the flat space Hagedorn temperature $\beta_H$, and it does not exist for lower temperatures.
	Numerical extrapolation of its thermodynamics properties shows a qualitative agreement with the small AdS black hole (see section \ref{sec:numerical_res}). This is a version of the correspondence principle that exists only in AdS.

	It is useful to understand the phase diagram of figure \ref{fig:phase_map} in terms of the holographic dual CFT.
	The thermal gas phase in the bulk is associated with the confined phase in the CFT, and the large AdS black hole with the deconfined phase \cite{Witten:1998qj,Witten:1998zw}.
	In the CFT at zero coupling, the two phases meet at the Hagedorn temperature without any further metastable phase, such as the small AdS black hole \cite{Sundborg:1999ue,Aharony:2003sx}. 
	At weak coupling there is an intermediate phase with an eigenvalue distribution that is non-uniform but ungapped, and higher loop computations are required to see if it is thermodynamically dominant~\cite{Aharony:2003sx}. One can conjecture \cite{Alvarez-Gaume:2005dvb} that this phase is continuously connected to a string star.
	In any case at strong coupling it is natural to expect the
	small AdS black hole phase to connect the large AdS black hole (deconfined) phase at high temperatures to the gas (confined) phase at low temperatures \cite{Aharony:2003sx}. Here we follow the speculation that this phase turns into a string star around $\beta \sim l_s$ before joining the confined phase, and study the latter behavior. In those terms, this work offers predictions on the connection of this metastable phase to the confined phase at strong coupling. Besides the value of the critical temperature $\beta_c$, we also find that the free energy close to $\beta_c$ is $F \sim (\beta-\beta_c)^2$ (see section \ref{sec:gas_transition}). This is the same scaling found in weak coupling \cite{Aharony:2003sx}, and it was expected on general grounds also for strong coupling \cite{Alvarez-Gaume:2005dvb}.

	\bigskip

	% Several open directions are possible. 
	% One may try to consider string star solutions with various charges and angular momenta.
	Even if the string star and the black hole are two continuously connected phases, a high order phase transition can still occur between them.
	In \cite{Chen:2021dsw} it was argued, using worldsheet methods, that in flat space type IIB a phase transition between the string star and the black hole is necessary. It would be interesting to see if similar arguments can be used directly in AdS as well.
	We can also consider these saddles in type IIB on $AdS_5\times S^5$ using the analysis of this work. The transition between the two happens in our setting when the two phases are ignorant of the AdS geometry, and thus the argument from \cite{Chen:2021dsw} follows. This theory is holographic to $\mathcal{N}=4$ super Yang-Mills, in which a third order Gross-Wadia-Witten phase transition was shown for the metastable phase in weak coupling \cite{Aharony:2003sx}.
	If this phase transition follows to strong coupling, as conjectured by \cite{Alvarez-Gaume:2005dvb,Alvarez-Gaume:2006fwd}, it might be the same phase transition advocated in \cite{Chen:2021dsw}.

	More broadly, one may hope to find the states corresponding to the AdS string star in the holographic CFT.
	At this point it seems very hard, first of all because the holographic CFT is strongly coupled. Also, this phase has no clear order parameter that distinguishes it from any other generic state at those temperatures. As a first step, further work can find the (qualitative) expectation value of a Polyakov loop (as a function of the temperature $\beta_\text{CFT}$) in this phase compared to the other Euclidean saddles.

\section{General properties of the AdS string star} \label{sec:properties}
	\subsection{The effective action}\label{sec:action}
	We are interested in the string theory Euclidean partition function on asymptotically $EAdS_{d+1}\times X_{9-d}$ with a conformal boundary of $S^{d-1}\times S^1$, where $X_{9-d}$ is some $9-d$ dimensional compact Euclidean manifold. 
	We will mostly focus on classical solutions in which the $X_{9-d}$ factorizes. As a result, we will consider the low energy effective action on (asymptotically) $EAdS_{d+1}$. Specifically, we will consider a stringy excitation around the thermal AdS geometry. It is given by the metric \cite{Witten:1998qj}
	\begin{equation}\label{eq:thermal_ads}
		ds^2 = \beta^2 \cosh^2(\rho/l_{ads}) dt^2 + d\rho^2 + l_{ads}^2\sinh^2(\rho/l_{ads}) d\Omega^2_{d-1},
	\end{equation}
	with $l_{ads}$ the AdS curvature scale. We identify the thermal circle by $t\sim t+1$ and use a dimensionfull temperature $\beta$ (related to the CFT dimensionless temperature by $\beta = l_{ads} \cdot \beta_\text{CFT}$).
	The topology of the solution is $R^d\times S^1$. We denote the $S^1$ radius at the origin ($\rho=0$) by $R_0=\beta/{2\pi}$. Notice that in AdS units, it is also the holographic CFT temporal radius. At any other point we have a radius $R(\rho) = \frac{1}{2\pi} \sqrt{g_{tt}} = \frac{\beta}{2\pi} \cosh(\rho/l_{ads})$. As $\rho$ increases the radius diverges exponentially. At a given $\rho$ we can consider the mass of a string winding around the $S^1$. At leading order in $l_s/l_{ads}$ the mass of such a mode is given by the flat space computation on $S^1\times R^d$, that we denote $m^2(R(\rho))$. At the (flat space) Hagedorn temperature $R_H=\beta_H/(2\pi)$ (that depends on the string theory in question) the winding mode becomes massless: $m^2(R=R_H)=0$. For example if we take $R_0=R_H$, the winding mode is massless at the origin $\rho=0$ of the thermal AdS, but exponentially massive as we increase $\rho$. For high-enough temperatures we can consider the effective theory on spatial slice of \eqref{eq:thermal_ads}. When $(R_0-R_H) \ll l_s$ the winding mode of the string is light enough deep inside the thermal AdS, that we can consistently add it to the effective $d$ dimensional theory. In this setting, we are looking for bound-state solutions for the winding mode and their thermal properties.

	The effective $d$ dimensional action is derived in appendix \ref{app:EFT}. We will consider solutions with constant $d$ dimensional dilaton $\phi_{d}=\text{const}$, and allow a gravitational back-reaction only of the $g_{tt}$ component. In other words, we consider the metric
	\begin{equation}\label{eq:thermal_fluct}
		ds^2 = \beta^2\cosh^2(\rho/l_{ads}) e^{2\varphi} dt^2 + d\rho^2 + \sinh^2(\rho/l_{ads}) d\Omega_{d-1}^2,
	\end{equation}
	where $\varphi$ describes the $U(1)$-invariant fluctuations of the thermal circle. We note that the $d+1$ dimensional dilaton is given by $\Phi = \phi_d + \varphi/2$, and as a result not a constant (see appendix \ref{app:EFT}).
	This approximation is valid when derivatives are small, so we will need to make sure our solution varies slowly enough. 

	We denote the string winding mode by $\chi(x)$, $x$ being the collective $d$ dimensional coordinate of $\rho,\Omega$. The resulting $d$ dimensional action is 
	\begin{equation}\label{eq:action_d}
	\begin{split}
		I_d &= \frac{1}{16\pi G_N} \int d^d x \sqrt{g^{(d)}} \, (2\pi R)\; e^{-2\phi_{d}} \\
		&\cdot \left( -\mathcal{R}+2\Lambda -4 (\nabla \phi_{d})^2 +  (\nabla \varphi)^2 
		+ |\nabla \chi|^2 + \left(m^2(R) + R \frac{\partial m^2}{\partial R} \varphi \right)|\chi|^2 + O({\alpha'})\right),
	\end{split}
	\end{equation}
	with the $d+1$ dimensional gravitational constant $G_N \simeq l_s^{d-1} g^2$ ($g$ is the $d+1$ dimensional string constant), $R(\rho)$ as given above, and the cosmological constant $\Lambda = -d(d-1)/(2 l_{ads}^2)$. 
	In \eqref{eq:action_d} we took only the leading order interaction between the winding mode $\chi$ and the metric mode $\varphi$. We also suppressed higher order interactions of $\chi$ with itself and with the metric. These approximations are valid when the amplitudes of $|\chi|,|\varphi|\ll 1$ are small. We will justify it a posteriori below. 
	Higher derivative and curvature terms are also suppressed. This is justified as long as the solution size is larger than the string scale $L \gg l_s$.

	For heterotic and type II string theories the winding mode mass and the (flat space) Hagedorn temperature are
	\begin{equation}\label{eq:m2_def}
	\begin{split}
		m^2_\text{Heterotic}(R) = \frac{R^2}{l_s^4}+\frac{1}{4R^2}
		-\frac{R^2_H}{l_s^4}-\frac{1}{4R^2_H}, \qquad & R_H^\text{Heterotic}/l_s = 1+\frac{1}{\sqrt{2}},\\
		m^2_\text{Type II}(R) = \frac{R^2-R^2_H}{l_s^4},\qquad &R_H^\text{Type II}/l_s = \sqrt{2}.
	\end{split}
	\end{equation}
	It will also be useful to define $\kappa \equiv \alpha' R_H \partial_R m^2(R_H)$. For heterotic and type II string theories
	\begin{equation} \label{eq:kappa_def}
		\kappa^\text{Heterotic} = 4\sqrt{2}, \qquad 
		\kappa^\text{Type II}=4.
	\end{equation}
	We stress that unlike in flat space, here $m^2(R)$ and $R\partial_R m^2(R)$ depend on the coordinate $\rho$ implicitly through $R(\rho)$. In order to consistently include $\chi$ in the action, we assumed its mass $m^2(R)\ll 1/l_s^2$ everywhere $\chi(\rho)$ is finite. At $\rho=0$ the condition on the temperature is $(R_0-R_H)/R_H \ll 1$. In particular notice that temperatures around $\beta \sim l_s$ are outside of this region.
	For $\rho \sim l_{ads}$ the approximation breaks for every temperature. If the solution length scale $L \ll l_{ads}$ then the error is exponentially suppressed. We will see below that the solution dynamically satisfies this condition, intuitively because the mass itself determines the exponential decay of the solution.

	\subsection{The equation of motion}\label{sec:EOM}
	Without loss of generality we take the constant value of the dilaton to be $\phi_d=0$ (otherwise it can be swallowed in the definition of $G_N$) and the metric \eqref{eq:thermal_fluct}. The equations of motion for the remaining $\varphi(x),\chi(x)$ fields are
	\begin{equation}\label{eq:eom_cov}
	\begin{split}
		\frac{1}{R} \nabla\cdot(R \vec\nabla\varphi) - \frac{1}{2} R \partial_R \; m^2 |\chi|^2 &= 0\\
		\frac{1}{R} \nabla\cdot(R \vec\nabla\chi) - m^2 \chi - R \partial_R m^2 \; \chi \varphi &=0
	\end{split}
	\end{equation}
	Here $\nabla\cdot,\vec\nabla$ are covariant derivatives in terms of the spatial $d$ dimensional slice. The first term on both lines can be understood as the thermal $AdS_{d+1}$ Laplacian acting on a $t$-invariant function.
	We will further assume that $\varphi,\chi$ are spherically symmetric and thus depend only on the radial coordinate $\rho$. The equations simplify to
	\begin{equation}\label{eq:eom_rho}
	\begin{split}
		\varphi''(\rho) &+ v(\rho) \cdot \varphi'(\rho) - \frac{1}{2} \left(R \partial_R m^2\right)(\rho) \cdot |\chi(\rho)|^2 = 0,\\
		\chi''(\rho) &+ v(\rho) \cdot  \chi'(\rho) - m^2(\rho) \cdot \chi(\rho) - \left(R \partial_R m^2\right)(\rho) \cdot \chi(\rho) \cdot \varphi(\rho) = 0,
	\end{split}
	\end{equation}
	with 
	\begin{align}
		R(\rho) &= R_0 \cosh(\rho/l_{ads}),\label{eq:R_def}\\
		v(\rho) &= \frac{1}{l_{ads}} \left(\tanh(\rho/l_{ads})+(d-1)\coth(\rho/l_{ads})\right),\label{eq:v_def}
	\end{align}
	and $m^2(R)$ defined in \eqref{eq:m2_def}.
	We can formally solve the first equation using an appropriate kernel
	\begin{equation}\label{eq:spherical_varphi_sol}
		\varphi(\rho) = l_{ads} \int_0^\infty d\rho' \; k(\rho/l_{ads},\rho'/l_{ads}) \left( \frac{1}{2} R \frac{\partial m^2}{\partial R} |\chi|^2\right)(\rho').
	\end{equation}
	The function $k(r,r')$ is defined in \eqref{eq:k_def}. In this form, we can write a single equation for $\chi(\rho)$
	\begin{equation}\label{eq:spherical_EOM}
	\begin{split}
		\chi''(\rho) &+ v(\rho) \cdot \chi'(\rho) - m^2(\rho) \cdot \chi(\rho) \\
		&- \frac{l_{ads}}{2} \left(R \partial_R m^2 \right)(\rho)\cdot \chi(\rho) \cdot
		\int_0^\infty d\rho' \; k\left(\rho/l_{ads},\rho'/l_{ads}\right) \cdot
		\left( R \partial_R m^2\right)(\rho') \cdot |\chi|^2(\rho')=0
	\end{split}
	\end{equation}

	\subsection{Thermodynamics}\label{sec:thermo}
	The string star is a classical solution of the effective gravitational action. As such, has free energy and entropy of order $1/G_N$, just like a black hole. In this section we will give the formal expressions of the string star thermodynamic properties. For comparison, at the end of the section we will describe the thermodynamic properties of other saddles that contribute to the same string theory partition function.
	
	We start by considering the thermodynamics of the string star saddle around thermal AdS. The free energy of the classical solution is its on-shell action \eqref{eq:action_d} divided by $\beta$. We will work with a scheme in which thermal AdS has zero on-shell action. We are thus left only with the contribution from $\chi,\varphi$:
	\begin{equation}
	\begin{split}
		F &= \frac{1}{16\pi G_N \beta} \int d^d x \sqrt{g^{(d)}} \, (2\pi R) \Bigg( 
		(\nabla \varphi)^2 + |\nabla \chi|^2 + \left(m^2 + R \partial_R m^2 \varphi \right)|\chi|^2	\Bigg).
	\end{split}
	\end{equation}
	It is convenient to define a normalized dimensionless free-energy $f$ by
	\begin{equation}
		F \equiv \frac{l_s^{d-2} \omega_{d-1}}{16\pi G_N} \; f,
	\end{equation}
	where $\omega_{d-1}$ the area of the unit ($d-1$)-sphere.
	Explicitly for the radial solution, the normalized free energy is
	\begin{equation}\label{eq:free_energy}
	\begin{split}
		f &= l_s^{2-d} \cdot \int_0^\infty d\rho V(\rho) \left( 
		(\varphi'(\rho))^2 + |\chi'(\rho)|^2 + \left(m^2(\rho) + R(\rho) \partial_R m^2(\rho) \varphi(\rho) \right) |\chi|^2(\rho)	\right),
	\end{split}
	\end{equation}
	with the radial volume
	\begin{equation}\label{eq:V_def}
		V(\rho) = l_{ads}^{d-1} \cosh(\rho/l_{ads}) \sinh^{d-1}(\rho/l_{ads}).
	\end{equation}

	The find the entropy of the classical solution, we take $\beta\partial_\beta-1=R_0 \partial_{R_0} -1$ of the action \eqref{eq:action_d}. Because $R(\rho)$ is exactly linear in $R_0$, we have
	\begin{equation}
	\begin{split}
		S & = (\beta \partial_{\beta} -1)I_d = \frac{1}{16\pi G_N} \int d^d x \sqrt{g^{(d)}} (2\pi R) \, R_0 \partial_{R_0} \mathcal{L}_d.
	\end{split}
	\end{equation}
	But because it is a classical action, the implicit $\partial_{R_0}$ is zero. We still have an explicit dependence on $R_0$ through $m^2$:
	\begin{equation}
		S = \frac{1}{16\pi G_N} \int d^d x \sqrt{g^{(d)}} (2\pi R) \left( (R\partial_R m^2) + ((R\partial_R)^2m^2) \; \varphi \right) |\chi|^2.
	\end{equation}
	It is similarly convenient to defined a dimensionless normalized entropy by
	\begin{equation}
		S \equiv \frac{l_s^{d-1} \omega_{d-1}}{16\pi G_N} \; s.
	\end{equation}
	Explicitly, the normalized entropy is given by the integral
	\begin{equation}\label{eq:entropy_def}
	\begin{split}
		s &= l_s^{1-d} \; \beta \cdot \int_0^\infty d\rho V(\rho) \left( (R\partial_R m^2)(\rho) + ((R\partial_R)^2m^2)(\rho) \; \varphi(\rho) \right)|\chi|^2(\rho).
	\end{split}
	\end{equation}

	\bigbreak
	
	We now turn to compare the string star's free energy to other Euclidean saddles. The results of the analysis are drawn schematically in figure \ref{fig:phase_map}. In the limit we are working with $|\chi|,|\varphi|\ll 1$ and so the string star free energy \eqref{eq:free_energy} is positive. The geometry of thermal AdS itself, also called the `gas phase', was chosen to have $F=0$ (at order $1/G_N$) and therefore the string star is subdominant by comparison.

	The AdS black hole is the $d+1$ dimensional Schwarzschild solution with negative cosmological constant $\Lambda$. Its horizon radius $r_h$ is related to the (asymptotic) temperature $\beta$ by
	\begin{equation}\label{eq:horizon_radius}
		\beta = \frac{4\pi \; l_{ads}^2 r_h}{d \; r_h^2 + (d-2) l_{ads}^2}.
	\end{equation}
	In terms of $r_h, \beta$, the black hole free energy and entropy are
	\begin{equation}
		F = \frac{\omega_{d-1}}{16\pi G_N} r_h^{d-2}\left(1-\frac{r_h^2}{l_{ads}^2}\right),\qquad S=\frac{\omega_{d-1}}{4G_N} r_h^{d-1}.
	\end{equation}	
	By \eqref{eq:horizon_radius}, AdS black holes exist only above a temperature $\beta < 2\pi/\sqrt{d(d-2)} \cdot l_{ads}$. For each such a temperature two possible horizon radii $r_h$ are possible (see figure \ref{fig:phase_map}). One branch of solutions is called `large AdS black holes'. At high temperatures $\beta \ll l_{ads}$ its horizon is $r_h = (4\pi/d) \cdot l_{ads}^2 / \beta$. The free energy and the entropy are (at leading order in $\beta/l_{ads}$)
	\begin{equation}
		F_\text{Large} = -\frac{\omega_{d-1}}{16\pi G_N} \left(\frac{4\pi}{d}\right)^d l_{ads}^{2d-2} \beta^{-d}, \qquad 
		S_\text{Large} = \frac{\omega_{d-1}}{G_N} \left(\frac{4\pi}{d}\right)^{d-1} l_{ads}^{2d-2} \beta^{1-d}.
	\end{equation}
	The large AdS black hole is a valid solution also around (and above) the Hagedorn temperature, as its horizon size there is huge. The free energy of the saddle is negative, and it is the dominant phase for high enough temperatures \cite{Hawking:1982dh,Witten:1998qj,Aharony:2019vgs}. In particular it is more dominant than the string star saddle whenever the latter is well defined. The second branch is termed the `small AdS black hole'. For temperatures $l_s \ll \beta \ll l_{ads}$ we find a horizon at $r_h = (d-2)/(4\pi) \cdot \beta$, and the solution is for small $\rho$ approximately an asymptotically flat Schwarzschild solution. As such, its free energy and entropy are given by
	\begin{equation}\label{eq:small_BH_S}
		F_\text{Small} = \frac{\omega_{d-1}}{16\pi G_N} \left(\frac{d-2}{4\pi}\right)^{d-2} \beta^{d-2}, \qquad 
		S_\text{Small} = \frac{\omega_{d-1}}{4 G_N} \left(\frac{d-2}{4\pi}\right)^{d-1} \beta^{d-1}.
	\end{equation}
	The small AdS black hole solution can't be trusted when $r_h \gtrsim l_s$ or equivalently $\beta \gtrsim l_s$. The string star solution is consistent for near-Hagedorn temperatures $(R_0-R_H)/R_H \ll 1$, and also fails around $\beta \sim l_s$ (see above). Therefore the two solutions share no regime of validity, and both break around $\beta \sim l_s$. For flat space, it was proposed that the two Euclidean solutions are non-perturbatively (in $\alpha'$) connected \cite{Horowitz:1997jc,Chen:2021dsw} (see also figure \ref{fig:length_scale}). We will see that in this intermediate regime both solutions are similar to flat space solutions, and so the question of connecting them in AdS is the same as in flat space.

	We can also consider the full $10$ dimensional geometry, which we assumed to be asymptotically $AdS_{d+1}\times X_{9-d}$. $10$ dimensional Schwarzschild solutions exist as long as their horizon is small compared to the curvature scale of $X_{9-d}$. As the latter is usually of the same order as $l_{ads}$, to leading order these are asymptotically flat $10$ dimensional Schwarzschild solutions. The $10$ dimensional gravitational constant is of order $G_N^{(10)} \sim G_N l_{ads}^{9-d}$ and so
	\begin{equation}
		F_\text{10d} = \frac{\omega_{8}}{16\pi G_N^{(10)}} \left(\frac{7 \beta}{4\pi} \right)^{7} \sim \frac{1}{G_N} l_{ads}^{d-9} \beta^{7}, \qquad 
		S_\text{10d} = \frac{\omega_{8}}{4\pi G_N^{(10)}} \left(\frac{7 \beta}{4\pi} \right)^{8}\sim \frac{1}{G_N} l_{ads}^{d-9} \beta^8.
	\end{equation}
	Just like the small AdS black hole, the solution is perturbative only for $\beta \gtrsim l_s$. In this regime we can compare its dominance to the other black holes we found. As the free energy is positive $F_\text{10d}>0$, it is subdominant to both the gas and the large AdS black hole. For $\beta \ll l_{ads}$ the solution has significantly lower free energy over the small AdS black holes, and is therefore more dominant. In section \ref{sec:therm_inst} we expand further on the relation between these two phases.

\section{Small solutions}\label{sec:small_sol}
	The lowest free energy solution of \eqref{eq:eom_rho} is a real monotonic profile for $\chi,\varphi$ that decays to zero for large enough $\rho$. We denote the length scale of the decay by $L$ (it is a function of the temperature $R_0$). 
	By ``small solutions" we mean solutions that decay fast enough so that the solution can be approximated by the flat space solutions found in \cite{Horowitz:1997jc}.
	In other words, we would like to consistently approximate \eqref{eq:eom_rho} by
	\begin{equation}\label{eq:flat_approx}
	\begin{split}
		R(\rho) &= R_0 +O(\rho^2/l_{ads}^2),\\
		v(\rho) &= \frac{d-1}{\rho}\left(1+O(\rho^2/l_{ads}^2)\right),\\
		m^2(\rho) &= m^2(R_0) + O(\rho^2/l^2_{ads}).
	\end{split}
	\end{equation}
	The resulting equations are exactly the equations studied by Horowitz and Polchinski \cite{Horowitz:1997jc} in the context of flat $R^d\times S^1$ (Note that we are not taking the flat space limit of AdS, but only considering small enough profiles that doesn't sense the curvature). They found that such solutions exist for $3\le d\le 5$. The amplitude of the solution scales like $|\chi|,|\varphi|\sim (R_0-R_H)/R_H$. As explained above, for the winding mode mass to be below the string scale we need $(R_0-R_H)/R_H \ll 1$. In this regime the amplitudes are small $|\chi|,|\varphi|\ll 1$. This is important to justify our earlier approximation of the action where we ignored higher order terms.

	The length scale of the solution $L$ grows with the temperature and is given by (see figure \ref{fig:length_scale})
	\begin{equation}\label{eq:flat_lengthscale}
		L/l_s \sim \left(\frac{R_0-R_H}{R_H}\right)^{-\frac{1}{2}}.
	\end{equation}
	In the regime $(R_0-R_H)/R_H \ll 1$ we have $L \gg l_s$, consistent with the suppression of higher derivative terms in the effective action.
	When the temperature is high enough, the solution is so large that the approximations \eqref{eq:flat_approx} are no longer valid.
	The leading correction is coming from the mass term, at order $\sim L^2/(l_s l_{ads})^2$. The solution is self-consistent as long as this correction is small compared to the leading mass term $m^2(R_0)$. The resulting condition on the length scale is $L^2 \ll l_s l_{ads}$, and on the temperature $(R_0-R_H)/R_H \gg l_s/l_{ads}$. Together, we trust the flat space solution only for temperatures
	\begin{equation}\label{eq:believe_flat}
		l_s/l_{ads} \ll \frac{R_0-R_H}{R_H} \ll 1.
	\end{equation}
	As we originally assumed a large gap $l_s/l_{ads}\ll 1$ in order to write the effective action, this regime is not empty.
	Finally, the entropy of the flat space solution was shown to be
	\begin{equation}\label{eq:flat_entropy}
		S \sim \frac{l_s^{d-1}}{G_N} \cdot \left( \frac{R_0-R_H}{R_H} \right)^{\frac{4-d}{2}}.
	\end{equation}

	What are the properties of the solution at the two ends of the validity region \eqref{eq:believe_flat}?
	At the low temperature limit $(R_0-R_H)/R_H \sim 1$ the length scale \eqref{eq:flat_lengthscale} is $L\sim l_s$, and the entropy \eqref{eq:flat_entropy} is $S \sim l_s^{d-1}/G_N$. 
	This is in qualitative agreement with a small AdS black hole of horizon size $r_h \sim l_s$, see \eqref{eq:small_BH_S}. This qualitative agreement between the two saddles is completely equivalent to the one already found for flat space, as both sides are much smaller than the AdS scale.
	As we said, in the near-Hagedorn limit $(R_0-R_H)/R_H \sim l_{s}/l_{ads}$ we have $L \sim \sqrt{l_s l_{ads}}$, with entropy $S\sim l_s^{d-1}/G_N \cdot (l_s/l_{ads})^{2-d/2}$.

	As in \cite{Horowitz:1997jc}, the solutions described in this section exist only for $3\le d \le 5$. In \cite{Chen:2021dsw} it was shown that above the flat space Hagedorn temperature no such solution are expected to exist for $d\ge 6$. By similar argument, we believe no such solutions exist in the temperatures range \eqref{eq:believe_flat} on AdS with $d\ge 6$. In practice, we didn't manage to find numerical solutions in this range (see section \ref{sec:numerical_res}).

\section{At the flat space Hagedorn temperature}\label{sec:at_hagedorn}
	We couldn't find a full analytical treatment for the solution at temperatures higher than the flat space regime \eqref{eq:believe_flat}. 
	As we review in section \ref{sec:numerical_res} below, numerically we find that solutions exist also around $R_H$. The solutions always have a length scale $L\sim \sqrt{l_s l_{ads}}$, and they merge with the (trivial) thermal AdS saddle for some $R_c < R_H$. 
	Here and in the next section we will describe analytical results that support this picture. In this section we study the solution exactly at the flat space Hagedorn temperature $R_0 = R_H$. At $R_0=R_H$, the leading correction to the winding mode mass around $\rho \approx 0$ is 
	\begin{equation}
		m^2(R_0=R_H) = \frac{\kappa}{2} \frac{\rho^2}{l_{ads}^2 l_s^2} +O(\rho^4/l_{ads}^4).
	\end{equation}
	Assuming the length scale of the solution is small enough ($L \ll l_{ads}$) for the higher orders above to be subleading, we can approximate \eqref{eq:eom_rho} by
	\begin{equation}\label{eq:eom_hag}
	\begin{split}
		\varphi''(\rho) & + \frac{d-1}{\rho} \varphi'(\rho) - \frac{\kappa}{2 \alpha'} \cdot |\chi(\rho)|^2 = 0,\\
		\chi''(\rho) &+ \frac{d-1}{\rho}  \chi'(\rho) - \frac{\kappa}{2} \frac{\rho^2}{l_{ads}^2 l_s^2} \chi(\rho) - \frac{\kappa}{\alpha'} \; \chi(\rho) \varphi(\rho) = 0.
	\end{split}
	\end{equation}
	Following \cite{Chen:2021dsw}, we can find a normalization in which the equations are parameter-independent. Taking
	\begin{equation}
		\varphi(\rho) = \frac{\alpha'}{\kappa L_H^2} \hat \varphi(\rho/L_H), \qquad \chi(\rho) = \frac{\alpha'}{\kappa L_H^2} \hat \chi(\rho/L_H),
	\end{equation}
	with 
	\begin{equation}\label{eq:L_H}
		L_H^2 = \frac{1}{\sqrt{\kappa}} l_s l_{ads}
	\end{equation}
	gives the following parameter-independent equations for $\hat\varphi(x),\hat\chi(x)$
	\begin{equation}\label{eq:hated_eq}
	\begin{split}
		\hat\varphi''(x) & + \frac{d-1}{x} \hat\varphi'(x) - \frac{1}{2} \cdot |\hat\chi(x)|^2 = 0,\\
		\hat\chi''(x) &+ \frac{d-1}{x}  \hat\chi'(x) - \frac{x^2}{2} \hat\chi(x) -  \; \hat\chi(x) \hat\varphi(x) = 0.
	\end{split}
	\end{equation}
	The solution for these equations are some $O(1)$ normalizable functions (notice that unlike flat space, here we have an $x^2$ term that serve as a potential). Therefore the amplitudes of the original variables are of order $|\chi|,|\varphi|\sim l_s/l_{ads} \ll 1$, which justifies our approximation to ignore higher order terms in the effective action. We also learn that the length scale of the $R_0=R_H$ solution is $L_H \sim \sqrt{l_s l_{ads}}$ \eqref{eq:L_H}. This is the same scale we got at the high temperature end of where we trusted the flat space solutions \eqref{eq:flat_lengthscale}. 
	Notice that $L_H \ll l_{ads}$ and so the approximation of \eqref{eq:eom_hag} is also justified. We can also approximate the entropy using \eqref{eq:entropy_def} to be
	\begin{equation}
		S\mid_{R_0=R_H} \approx \frac{\zeta}{16\pi G_N} \; \frac{2\pi R_H \; l_s^{d-2}}{\kappa^{\frac{d}{4}}} \; \left(\frac{l_s}{l_{ads}}\right)^\frac{4-d}{2}.
	\end{equation}
	Here $\zeta$ is an $O(1)$ dimensionless number, defined as the $R^d$ norm of the solution to \eqref{eq:hated_eq}:
	\begin{equation}
		\zeta = \int d^d x |\hat \chi(x)|^2.
	\end{equation}
	This is of the same order as the entropy we got at the high temperature end in which we trust the flat space entropy \eqref{eq:flat_entropy}. It thus seems like nothing drastic is happening close to $R_H$ for $3\le d \le 5$, and the solution is qualitatively similar to the solutions around $(R_0-R_H)/R_H \sim l_s/l_{ads}$ (see figure \ref{fig:length_scale}).

\section{Evaporation to gas}\label{sec:gas_transition}
	We now turn to study the solutions for $R_0<R_H$. For these temperatures the mass squared at the origin $\rho=0$, which we denote by $m^2_0\equiv m^2(R_0)$, is negative.
	Close enough to $\rho=0$, the solution of \eqref{eq:eom_rho} with $m^2_0 < 0$ will oscillate. For large enough $\rho$ the mass squared turns positive, and grows exponentially with $\rho$. At large $\rho \gg l_{ads}$ we have an exploding and a decaying mode.
	We therefore expect that a discrete set of normalizable solutions exists, one for each number of oscillations in the small $\rho$ region. Of those solutions only the first one is in our interest, as it is connected continuously to the solutions we found for higher $R_0$. The others are highly unstable solutions that won't concern us. 
	We further expect that above some critical temperature $R_0 = R_c < R_H$ this solution will cease to exist. We will assume the simplest scenario: that at the critical temperature $R_0=R_c$ the solution merges with the trivial $\chi(\rho)=\varphi(\rho)=0$ solution (see figure \ref{fig:phase_map}). This assumption is justified by the numerical evidence, see section \ref{sec:numerical_res}.
	As a result we will be able to find $R_c$ and estimate the thermodynamic behavior at that point.

	% For brevity denote the equations of motion \eqref{eq:eom_rho} as a function of $R_0$ by $F[f(\rho),{R_0}]=0$. Here $f(\rho)$ stands collectively for both $\varphi(\rho)$ and $\chi(\rho)$. The trivial solution $f(\rho)=0$ is a solution of \eqref{eq:eom_rho} for every $R_0$. We are interested in the properties of the non-trivial solution, which we denote $f_{{R_0}}(\rho)$. As both are solutions we have
	% \begin{equation} \label{eq:sols}
	% 	F[0,{R_0}] = F[f_{R_0}(\rho),{R_0}] = 0.
	% \end{equation}
	% At the critical value $R_0=R_c$ the non-trivial solution merges with the trivial one. We therefore assume that at leading order in $R_0-R_c$
	% \begin{equation}\label{eq:small_R_f}
	% 	f_{{R_0}}(\rho) = 0 + \delta f(\rho) (R_0-R_c)^\alpha +O(({R_0}-R_c)^{3\alpha}),
	% \end{equation}
	% for some functions $\delta f(\rho)$ and a positive power $\alpha>0$. Numerically we find $\alpha \approx 0.5$.
	% Deriving both sides of \eqref{eq:sols} by $R_0$ at ${R_0}=R_c$ gives
	% \begin{equation}
	% 	\frac{\partial F}{\partial {R_0}}[0,R_c] = 
	% 	\frac{\partial F}{\partial {R_0}}[0,R_c] + (R_0-R_c)^{\alpha-1} \int d\rho \frac{\delta F}{\delta f(\rho)}[0,R_c] \cdot \delta f(\rho) + O((R_0-R_c)^1),
	% \end{equation}
	% or (assuming further $\alpha\le 1$)
	% \begin{equation}
	% 	\int d\rho \frac{\delta F}{\delta f(\rho)}[0,R_c] \cdot \delta f(\rho) = 0.
	% \end{equation}
	If the solution coincides with the trivial solution $\varphi=\chi=0$ at $R_0=R_c$, it means that at that point the trivial solution has a zero mode.
	In other words, the linearized equations around the trivial solution should have a normalizable solution at $R_0=R_c$.  This is analogous to the flat space case, where at the Hagedorn temperature $R_H$ the condensate is massless. If we define the Hagedorn temperature as the the temperature above which the string gas phase becomes tachyonic, then $R_c$ is also the ``AdS Hagedorn temperature". $R_c$ should be understood as a correction of the flat space Hagedorn temperature $R_H$ due to the AdS curvature.
	Expanding \eqref{eq:eom_rho} to linear order around the trivial solution $\chi = 0+ \delta \chi$, $\varphi = 0+\delta\varphi$ gives the linearized (decoupled) equations
	\begin{equation}\label{eq:EOM_lin}
	\begin{split}
		\delta\varphi''(\rho) &+ v(\rho) \cdot \delta\varphi'(\rho) = 0,\\
		\delta\chi''(\rho) &+ v(\rho) \cdot \delta\chi'(\rho) - m^2(\rho) \delta\chi(\rho) = 0.
	\end{split}
	\end{equation}
	The equation for $\varphi$ is nothing but the free massless field in thermal AdS. There are no non-singular normalizable solutions, as the boundary conditions $\delta\varphi'(0)=\delta\varphi(\infty)=0$ leave only the trivial solution $\delta \varphi=0$.
	We will assume that $\delta \chi$ has a solution at $R_0=R_c$ with length scale $L_c$ that decays fast enough (and justify it at the end). We can approximate the linear equation for $\delta \chi$ up to sub-leading corrections in $\rho/l_{ads}$. By derivatives suppression, the first non-trivial term is coming solely from the mass term. In other words, we substitute in \eqref{eq:EOM_lin}
	\begin{equation}\label{eq:values}
	\begin{split}
		v(\rho) &= \frac{d-1}{\rho}+ O\left(\rho/l^2_{ads}\right),\\
		m^2(\rho) &=  m^2_c + \frac{\kappa_c}{2} \; \frac{\rho^2}{l_s^2 l_{ads}^2} + O\left(\rho^4/(l_{ads}^4 l_s^2)\right).
	\end{split}
	\end{equation}
	Here we mean $m^2_c=m^2(R_c)$ and $\kappa_c = \alpha' (R\partial_R m^2)(R_c)$. 
	The resulting differential equation can be solved analytically. To have a non-singular solution at $\rho=0$ we demand $\delta \chi'(0)=0$, and arbitrarily choose $\delta \chi(0)$. The solution is
	\begin{equation} \label{eq:approx_sol}
	\begin{split}
		\delta\chi(\rho) &= e^{-\frac{\rho^2}{2 L^2}}\cdot L_{-\nu}^{\left(\frac{d-2}{2}\right)}\left(\rho^2/L^2\right),\\
		L^2 &= \sqrt{\frac{2}{\kappa_c}} l_s l_{ads},\\
		\nu &= \frac{1}{4}\left(d+ L^2 \; m_c^2\right).
	\end{split}
	\end{equation}
	$L_{-\nu}^{(\alpha)}(x)$ is the generalized Laguerre polynomial. $L_{-\nu}^{(\alpha)}(x)$ is analytic around $x=0$ and generically exponentially diverges for large $x\gg 1$,
	\begin{equation}
		L_{-\nu}^{(\alpha)}(x) \sim \frac{e^{x} x^{\nu-1}}{\Gamma(\nu)}.
	\end{equation}
	As a result, the large $\rho$ behavior of the solution \eqref{eq:approx_sol} is diverging with
	\begin{equation}
		\delta \chi(\rho) \sim \frac{(\rho/L)^{2\nu-2}}{\Gamma(\nu)}
		e^{\frac{\rho^2}{2 L^2}}.
	\end{equation}
	The solution is non-normalizable as long as $1/\Gamma(\nu)\ne 0$. We learn that there exist normalizable solutions for non-negative integer $\nu \in {\mathbb Z}_0$. 

	Substituting \eqref{eq:values} in \eqref{eq:EOM_lin} gives a Hamiltonian-like equation for $\delta\chi(\rho)$. In this analogy the energy of the solution is parametrized by the temperature $R_c$ as $E = -m_c^2$. The potential is quadratic at this order and so we expect a discrete set of eigenfunctions for $\delta \chi(\rho)$. These are exactly the solutions we get by demanding (as a function of the temperature $R_0$) $\nu\in {\mathbb Z}_0$. We are interested in the ``ground-state wave-function" that corresponds to the $\nu=0$ solution. This is the zero mode associated with the merging of the full solution of \eqref{eq:eom_rho} to the trivial solution. Expanding the equation $\nu = 0$ to leading order in $l_s/l_{ads}$ gives the solution for the AdS Hagedorn temperature $R_c$. At leading order in $l_s/l_{ads}$ (in which we can take $\kappa_c\approx \kappa$) we get
	\begin{equation}\label{eq:R_c}
		R_c / R_H = 1 - \frac{d}{\sqrt{2 \kappa}} \cdot\frac{l_s}{l_{ads}}
		+ O\left(\frac{l_s^2}{l_{ads}^2}\right).
	\end{equation}
	As we explained above, $R_c$ is the AdS Hagedorn temperature above which the thermal gas becomes tachyonic. By the holographic dictionary it is related to the holographic CFT's Hagedorn temperature by $T_c^\text{CFT} = l_{ads}/\beta_c$.
	For type IIB on $AdS_5\times S^5$ we can write the result in terms of the dual $\mathcal{N}=4$ super Yang-Mills theory by $(l_{ads}/l_s)^4=\lambda$, where $\lambda \equiv g_{YM}^2 N$ is the CFT 't Hooft coupling, to get (using \eqref{eq:m2_def}, \eqref{eq:kappa_def})
	\begin{equation}
		T_c^\text{CFT} = \frac{1}{\sqrt{8 \pi^2}} \lambda^\frac{1}{4} + \frac{1}{2\pi} +O\left(\lambda^{-\frac{1}{4}}\right).
	\end{equation}
	This result was independently found by \cite{maldacena_private}.
	In \cite{Harmark:2021qma}, the CFT Hagedorn temperature in $\mathcal{N}=4$ was found using integrability methods numerically for every value of $\lambda$. A linear fit to the large $\lambda$ behavior gave the $\lambda^0$ coefficient  $c_1 \approx 0.159$, which agrees with our $1/(2\pi)=0.15915...$.\footnote{We thank J. Maldacena for mentioning the connection to \cite{Harmark:2021qma}.}
	In $AdS_3$ with NS-NS flux the Hagedorn temperature was exactly computed in 
	\cite{Lin:2007gi}, with a first correction to $R_c/R_H$ of order $l_s^2/l_{ads}^2$. This contradiction with \eqref{eq:R_c} is explained by the fact that the NS-NS flux affects the mass of the winding mode and modifies our computation. This does not happen for R-R flux.\footnote{We thank D. Kutasov for mentioning the connection to \cite{Lin:2007gi}.}

	Because $L_{0}^{(\alpha)}(x)=1$, the solution at this value is simply $\delta \chi(\rho) = e^{-\frac{\rho^2}{2L^2_c}}$. The critical length scale of the solution is
	\begin{equation}\label{eq:L_c}
		L_c^2 = \sqrt{\frac{2}{\kappa}} \; l_s l_{ads}.
	\end{equation}
	Notice that the flat space limit $l_{ads}=\infty$ gives an infinitely large solution, as we expect in flat space at the massless limit $R_c = R_H$. Also notice that $L_c \ll l_{ads}$ thus justifying our approximation \eqref{eq:values}. One can wonder whether higher curvature terms in the effective action can change \eqref{eq:R_c} at the same order. The leading (in orders of $1/l_{ads}$) possible term is $\mathcal{R} |\chi|^2$ which shifts the condensate mass by order $1/l_{ads}^2$. As a result the definition of $R_H$ is shifted by order $l_s^3/l_{ads}^2$. This in turn gives a subleading change to the value of $R_c$ \eqref{eq:R_c}, and so we can ignore it. Higher derivative corrections start at the same subleading order: the first $\alpha' \partial^4 |\chi|^2$ term also scales like $l_s^2/L_c^4 \sim 1/l^2_{ads}$.

	The results can be intuitively understood as follow. Expanding the mass \eqref{eq:values} around $R_c\approx R_H$ and small $\rho \ll l_{ads}$ gives
	\begin{equation}\label{eq:mass_approx}
		\alpha' m^2(\rho) \approx \kappa \frac{R_c-R_H}{R_H} + \frac{\kappa}{2} \; \frac{\rho^2}{l_{ads}^2} .
	\end{equation}
	Around $\rho=0$ the mass is negative and we can define $-k^2_c \equiv m^2_c = \kappa/l_s^2  \frac{R_c-R_H}{R_H}$. As a result we expect $\chi(\rho)$ to oscillate with frequency $k$ for small enough $\rho$. In order for the solution to stop oscillating after one turn, at the scale $\rho \sim 1/k_c$ the mass needs to vanish $m^2(\rho\sim 1/k_c)\sim 0$. Using \eqref{eq:mass_approx} we can solve the constraint and get $k_c^2 \sim 1/(l_s l_{ads})$ which agrees with \eqref{eq:L_c}. Using the relation between $k_c$ and the temperature, we also get an agreement with \eqref{eq:R_c}.

	\bigbreak

	The linear equation for $\delta \chi$ can't fix its value at the origin $\delta\chi(0)$. To find it, we need to go to the first non-linear order. For now we set $\delta \chi(\rho) = A \cdot \delta \hat \chi(\rho/L_c)$ with $\delta \hat \chi(x)=\exp(-x^2/2)$, and assume $A\ll 1$ (close to $R_0=R_c$). Our goal is to find $A$ close to $R_0=R_c$ using the expansion of \eqref{eq:eom_rho} in orders of $(R_0-R_c)$. At leading order in $A$, the equation for $\delta \varphi$ is
	\begin{equation}
		\delta\varphi''(\rho) + v(\rho) \cdot \delta\varphi'(\rho) = \frac{\kappa_c}{2 l_s^2} |\delta\chi|^2.
	\end{equation}
	The solution is unique and proportional to $A^2 \kappa_c (L_c/l_s)^2 $. So we define $\delta \varphi(\rho) = A^2 \kappa_c (L_c/l_s)^2  \cdot \delta\hat\varphi(\rho/L_c)$. For the next order we define $\chi(\rho) = \delta \chi(\rho) + \chi_{(2)}(\rho)+...$, where we assume that $\chi_{(2)}$ is subleading in $A$. To find the equation for $\chi_{(2)}$ we need to expand the equations \eqref{eq:eom_rho} to the next order in $R_0-R_c$. The result is
	\begin{equation}
		\chi_{(2)}''(\rho) + v(\rho) \cdot \chi_{(2)}'(\rho) 
		- \left(m^2_c + \frac{\kappa_c}{2} \; \frac{\rho^2}{l_s^2 l_{ads}^2}\right) \chi_{(2)}(\rho) = 
		\frac{\kappa_c}{l_s^2} \left(\frac{R_0-R_c}{R_c} + \delta\varphi(\rho)\right) \delta\chi(\rho).
	\end{equation}
	The LHS is the same linear equation we had for $\delta\chi$ above. The first term on the RHS is coming from the leading correction to the mass, and the second from the interaction term at leading order. The two terms have different scaling with $A$ and opposite signs (by \eqref{eq:spherical_varphi_sol} and \eqref{eq:k_def} $\delta \hat \varphi$ is negative). For $\chi_{(2)}$ to be normalizable, we need to tune the source and in this way find $A$. We didn't do it here, but by comparing the two terms the result is of order $A^2 \sim (l_s/l_{ads}) \cdot (R_0-R_c)/R_c$. In other words, close to $R_0 \approx R_c$ we expect
	\begin{equation}\label{eq:R_c_amplitude}
		|\chi| \sim \left(\frac{1}{\kappa}\frac{l_s}{l_{ads}} \frac{R_0-R_c}{R_c}\right)^{\frac{1}{2}}, \qquad |\varphi| \sim \frac{R_0-R_c}{R_c}.
	\end{equation}
	Note that the $R_0-R_c$ scaling of the $\varphi |\chi|^2$ interaction term is the same as that of the $|\chi|^4$ term~\cite{Atick:1988si,Dine:2003ca,Brustein:2021ifl}, which we ignored in the effective action above. But due to the $l_s/l_{ads}$ scaling we have $|\chi|^4 \ll \varphi |\chi|^2$, and the approximation is still consistent.

	In the previous section we saw that at $R_0=R_H$ the length scale of the solution was also $L_H \sim \sqrt{l_s l_{ads}}$ \eqref{eq:L_H}. The amplitudes we found \eqref{eq:R_c_amplitude} also coincide with the results at $R_0=R_H$. Setting the value of $R_c$ \eqref{eq:R_c} inside \eqref{eq:R_c_amplitude}, gives $|\chi|,|\varphi| \sim l_s/l_{ads}$ at $R_0=R_H$. It seems like for $3\le d \le 5$ no surprises happen between $R_H$ and $R_c$ (see also section \ref{sec:numerical_res}).
	As a result of \eqref{eq:R_c_amplitude}, the scaling of the free energy and the entropy close to $R_c$ are (up to order $1$ factors)
	\begin{equation}\label{eq:thermo_R_c}
	\begin{split}
		F \sim \frac{l_s^{d-2}}{G_N} \left(\frac{l_s}{l_{ads}}\right)^\frac{2-d}{2} \; \left(\frac{R_0-R_c}{R_c}\right)^2, \qquad 
		S \sim \frac{l_s^{d-1}}{G_N} \left(\frac{l_s}{l_{ads}}\right)^\frac{2-d}{2} \; \frac{R_0-R_c}{R_c}
	\end{split}
	\end{equation}
	The numerical evidence (see figure \ref{fig:entropy_num} below) seems to agree with this prediction. Note that the scaling of \eqref{eq:R_c_amplitude} and \eqref{eq:thermo_R_c} with $R_0-R_c$ has a simple argument. For any $R_0$, denote the amplitude of the normalizable eigenmode (of the quadratic theory) that will be the zero mode at $R_c$ by $A$. Close to $R_c$ its amplitude is small. The effective action for $A$ would be at leading order 
	\begin{equation}\label{eq:eft_A}
		I_\text{eff}(A) = a (R_0-R_c) A^2 + b A^4 + O(A^6),
	\end{equation}
	for some $a>0$ and $b<0$ (because $\delta\varphi<0$). Minimizing the action immediately gives \eqref{eq:R_c_amplitude} and \eqref{eq:thermo_R_c}.
	In terms of the hologrpahic CFT, these scalings agree with the predictions at weak coupling \cite{Aharony:2003sx}, where such an EFT was derived for the trace of the $S^1$ holonomy of the gauge field. The scaling \eqref{eq:thermo_R_c} was also advocated for the small AdS black hole phase in \cite{Alvarez-Gaume:2005dvb,Alvarez-Gaume:2006fwd}.

	The results of this section are valid for any spatial dimension $d$. It may come as a surprise, after all in flat space \cite{Horowitz:1997jc} and in AdS at lower temperatures (sections \ref{sec:small_sol} and \ref{sec:at_hagedorn}) solutions were found only for $3\le d \le 5$. Nevertheless the conclusion of this section is that the AdS string star exists for every $d$ as a reliable solution close enough to the AdS Hagedorn temperature $(R_0-R_c)/R_c \ll 1$ with \eqref{eq:thermo_R_c}. For $3\le d \le 5$ the string star persists as a solution to lower temperature until $(R_0-R_H)/R_0 \lesssim 1$. For higher dimensions $d\ge 6$ we believe the solution becomes unreliable (as $L \sim l_s$ and $|\chi|,|\varphi| \sim 1$) somewhere between $R_c < R_0 \le R_H$. As we show numerically in section \ref{sec:numerical_res}, the length scale and the normalized entropy are $O(1)$ at the low temperature end of the validity region, just like a string sized black hole. We found that in AdS a correspondence principle with the small AdS black hole is plausible also for $d\ge 6$, except that there the string star phase becomes reliable only above the flat space Hagedorn temperature $R_H$. In this context, \cite{Chen:2021emg} argued that for large enough $d$, string-size black hole solutions persist above the flat space Hagedorn temperature, which fits nicely with our proposed picture.

	Finally, we note that too close to $R_0=R_c$ we no longer trust the tree level calculation to represent the free energy of the phase. In this limit the tree level result vanishes and a normalizable mode turn massless. The 1-loop result thus becomes the leading contribution close enough to $R_c$. This an avatar of the string gas instability at the AdS Hagedorn temperature $R_c$. One way to get the condition on the temperature is to require the Euclidean action close to $R_c$ to be large $I_d \gg 1$. The result is
	\begin{equation}
	 	\frac{R_0-R_c}{R_c} \gg \left(\frac{l_s}{l_{ads}}\right)^{\frac{d-2}{4}} g,
	 \end{equation}
	where $g$ is the $d+1$ dimensional string coupling.

\section{Thermodynamic instabilities} \label{sec:therm_inst}
	In this section we will study the thermodynamic stability of the solutions we found. As we reviewed in section \ref{sec:thermo} above (and plotted in figure \ref{fig:phase_map}), the free energy of the string star is positive, and thus larger than both the gas (thermal AdS) and the large black hole saddles. This phase is therefore canonically unstable. As these phases are different Euclidean saddles for small $G_N$, this instability is non-perturbative in $G_N$.

	At higher temperatures we hope the string star is continuously connected to the Euclidean saddle of a small AdS black hole. The latter is a well defined Euclidean saddle only for temperatures $l_s \ll R_0$. Just like the string star solution, this phase is also subdominant compared to the large AdS black hole and the gas phases. But apart from these non-perturbative instabilities, the small AdS black hole also has a perturbative instability: the quadratic variation matrix of the solution's Euclidean action has a normalizable negative eigenmode~\cite{gross:1982}. The negative mode is related to the fact this phase also has negative specific heat. Negative specific heat is yet another sort of thermodynamic instability. But for black holes one can show it is actually related to the negative eigenmode~\cite{Reall:2001ag}. Intuitively the specific heat is related to the (quadratic) decrease of the free energy when changing the size of the black hole. The negative mode of \cite{gross:1982} exactly changes (off-shell) the size of the black hole on the spatial metric components. The decrease of the free-energy at second order by the temperature (due to negative specific heat) is then mapped to the second variation of this normalizable mode.

	Does the string star have such a negative mode as well? For the flat space string star it was shown in \cite{Horowitz:1997jc,Chen:2021dsw} that such a mode exists. As a result, such a mode exists also in AdS, whenever we trust the flat space approximation \eqref{eq:believe_flat}. We believe that such a negative mode exists also for higher temperatures, although we couldn't find a general argument. Somewhat similar to the black hole saddle, close to the AdS Hagedorn temperature $R_c$ we can use the specific heat to argue for the existence of such a negative mode. The string star solution to \eqref{eq:eom_rho} implicitly depends on $R_0$ through the equations' dependence on $R_0$. For each $R_0$ the asymptotic behavior of both $\varphi(\rho), \chi(\rho)$ remains the same: they decays exponentially (or faster than that) to zero.
	Our ansatz for a negative mode is the $R_0$ derivative of the solution itself:\footnote{We stress that both here and in the black hole case there is a mode that changes the asymptotic (holographic) temperature. The free-energy second variation of this mode is exactly the specific heat, but this mode is not a normalizable one, as it changes the boundary conditions of the metric. As such, this mode is not a valid perturbation, and not the one we describe.} 
	\begin{equation} \label{eq:negative_mode_ansatz}
		\delta \varphi(\rho)  = R_0 \partial_{R_0} \varphi(\rho), \qquad\delta \chi(\rho)  = R_0 \partial_{R_0} \chi(\rho).
	\end{equation}
	These modes are clearly normalizable. Taking the temperature derivative of \eqref{eq:eom_rho}, we can find an explicit expression for the second variation of the free energy \eqref{eq:free_energy} with \eqref{eq:negative_mode_ansatz}:
	\begin{equation}
	\begin{split}
		\delta^2 f = - l_s^{2-d} \cdot \int_0^\infty d\rho V(\rho) &\left( 
		  (R\partial_R)^2 m^2 \left(\frac{1}{2} \; |\chi|^2 \delta\varphi + \varphi \chi  \delta\chi \right) 
		  +R\partial_R m^2 \; \chi \delta\chi \right).
	\end{split}
	\end{equation}
	From this expression it is yet unclear if the variation is negative or not.
	At $R_0=R_c$ the second variation vanishes $\delta^2 f =0$, as at that temperature the solution is trivial $\chi=\varphi=0$.
	Analogously, at $R_0=R_c$ the mode \eqref{eq:negative_mode_ansatz} is exactly the zero mode \eqref{eq:approx_sol}. Therefore at $R_0=R_c$ the ansatz is not a negative mode but a zero mode.

	Taking the full $(R_0\partial_{R_0})^2$ operator on the normalized free energy \eqref{eq:free_energy}, we can relate the variation above (which corresponds to deriving only the solution) to the entropy and the specific heat:
	\begin{equation}\label{eq:d2f_expr}
	\begin{split}
	 	\delta^2 f &= \frac{l_s^{2-d}}{2} \int_0^\infty d\rho V(\rho) 
	 	\left\{ \left(R\partial_R + (R \partial_R)^2\right) m^2 
	 	+ \varphi(\rho) \cdot \left((R \partial_R)^2+(R \partial_R)^3\right) m^2\right\} |\chi|^2(\rho)\\
	 	& \quad - \frac{l_s}{2} \partial_{\beta} s.
	 \end{split}
	\end{equation}
	In the second line $s$ is the normalized entropy \eqref{eq:entropy_def}. The first term is due to the explicit dependence on $R_0$ inside the action, and has no dependence on the variation. To good approximation, one can consider only the first term inside the curly bracket, and get that this term is positive.
	The second term on the RHS is positively proportional to the specific heat $C = \frac{dS}{dT} = -\beta^2 \partial_\beta S$. 
	In order to show $\delta^2 f$ is negative, we need the specific heat term to be negative `enough'. Close to $R_c$ we found the entropy scales like $S\sim (R_0-R_c)^1$ \eqref{eq:thermo_R_c}. As a result the specific heat is negative and of order $(R_0-R_c)^0$. The integral in \eqref{eq:d2f_expr} on the other hand scales like $(R_0-R_c)^1$. Therefore close enough to $R_c$ \eqref{eq:negative_mode_ansatz} is indeed negative.

	What happens to this mode at lower temperatures? For $d=3$ and in the flat space regime the specific heat is still negative. In that case the integral in \eqref{eq:d2f_expr} is of order $(R_0-R_H)^{1/2}$. The specific heat on the other hand scales like $-(R_0-R_H)^{-1/2}$. As we have $(R_0-R_H)/R_H \ll 1$, the specific heat controls \eqref{eq:d2f_expr} and the mode is still negative $\delta^2 f <0$. For $d=4,5$ we don't expect \eqref{eq:negative_mode_ansatz} to be negative for lower temperatures as the specific heat is non negative. Nevertheless, we believe the negative mode we found close to $R_c$ is continuously connected (as a function of the temperature) to the negative mode found in \cite{Horowitz:1997jc,Chen:2021dsw}.	
	The existence of a negative mode at the two ends of the temperature range seems to suggest the string star saddle does have a negative mode in the entire range (for all $3\le d \le 5$).

	\bigbreak

	The last type of instability we will consider has to do with the full ten dimensional geometry. We assumed our full geometry is (asymptotically $EAdS_{d+1}$) $\times X_{9-d}$, where $X_{9-d}$ is compact with the characteristic size $\sim l_{ads}$ (the AdS curvature scale).
	Before explaining the instability of the string star phase, we turn again to the small AdS black hole saddle. This time, we start by considering the Lorentzian solution. In terms of the $9+1$ dimensional geometry, this solution is homogeneous along the $X_{9-d}$ directions.
	When the AdS black hole horizon is much smaller than the size of the compact direction $r_h \ll l_{ads}$ there's a classical mode of the spatial $d$ dimensional metric, modulated around the $X_{9-d}$ coordinates, that grows exponentially with time \cite{Hubeny:2002xn,Buchel:2015gxa,Buchel:2015pla}.
	This is a classical (or dynamical) instability, that breaks the homogeneity of the solution. It is known as the Gregory-Laflamme (GL) instability \cite{Gregory:1993vy,Kol:2004ww,Gregory:2011kh}. In this way small AdS black holes turn into $10$ dimensional black holes \cite{Dias:2015pda}. In other words, this is the dynamical manifestation of the $10$ dimensional black hole being thermodynamically favored over a small AdS black hole (see section \ref{sec:thermo} and figure \ref{fig:phase_map}) \cite{Banks:1998dd,Horowitz:1999uv,Prestidge:1999uq}. The relation between the two instabilities however is so far indirect, as we just described a Lorentzian and classical instability, and in thermodynamics we consider Euclidean saddles and the off-shell modes around them. Nevertheless the relation can be shown more rigorously. The dynamical modes are eventually equivalent to $10$ dimensional Euclidean negative modes, and both are related to the same negative mode of the ($d+1$ dimensional) small AdS black hole that we mentioned at the beginning of the section \cite{Gubser:2000ec,Gubser:2000mm,Prestidge:1999uq,Reall:2001ag,Hubeny:2002xn,Kol:2004ww}. 
	The basic idea is that for high enough frequency along $X_{9-d}$ the time exponent turns to zero. At that point we have a stationary solution that can be trivially Wick-rotated to a zero mode of the Euclidean $10$ dimensional solution. Thinking of the kinetic energy on $X_{9-d}$ as a linear source, this mode also describes a negative (off-shell) mode in $d+1$ Euclidean dimensions.

	As a solution in $9$ dimensions (after compactifying time), the string star saddle we considered in this work is homogeneous along the $X_{9-d}$ directions.
	We would like to ask whether, as in the small AdS black hole phase, the solution has a GL-like instabilities. As we need to compactify Euclidean time to describe this phase, it is unclear how to look for Lorentzian GL instabilities. In the Euclidean setting however, we can still ask whether there are negative $9$ dimensional modes of the solution. Similar to the small AdS black hole saddle, these modes are related to the $d$ dimensional negative mode discussed at the beginning of the section, as we will show now.

	First, and following the arguments above, we will assume a negative eigenmode of the $d$ dimensional solution exists. Denote this variation of the solution by $\delta \varphi(\rho)$ and $\delta \chi(\rho)$. We also denote the variation's eigenvalue with respect to the action's second variation by $\lambda<0$. The variation thus satisfies
	\begin{equation}\label{eq:thermo_instability}
	\begin{split}
		 \delta\varphi''+v(\rho) \delta\varphi' - R\partial_Rm^2 \; \chi \delta\chi &= -\lambda \; \delta \varphi,\\
		 \delta\chi''+v(\rho) \delta\chi'-m^2 \; \delta \chi - R\partial_R m^2 (\chi \delta\varphi + \varphi \delta \chi) &= -\lambda \; \delta\chi.
	\end{split}
	\end{equation}
	Next, we turn to describe the higher dimensional instability. For simplicity, we add a single non-compact direction to the metric \eqref{eq:thermal_ads}:
	\begin{equation}\label{eq:metric_with_z}
		ds^2 = \beta^2 \cosh^2(\rho/l_{ads}) dt^2 + d\rho^2 + l_{ads}^2\sinh^2(\rho/l_{ads}) d\Omega^2_{d-1} + dz^2.
	\end{equation}
	This is a good approximation assuming the wavelength of the mode (along $X_{9-d}$) we will consider is much smaller than the radius of $X_{9-d}$ (of order $l_{ads}$). The $z$-independent solution we described in the previous sections is of course still a solution. We can consider quadratic off-shell variations of this solutions that depend also on the $z$ coordinate which we denote $\delta \varphi(\rho,z)$, $\delta \chi(\rho,z)$. An eigenstate of the higher dimensional action's second variation will satisfy
	\begin{equation}\label{eq:thermo_instability_GL}
	\begin{split}
		 \partial_z^2\delta\varphi+\delta\varphi''+v(\rho) \delta\varphi' - R\partial_Rm^2 \; \chi \delta\chi &= -\Lambda \; \delta \varphi,\\
		 \partial_z^2\chi+\delta\chi''+v(\rho) \delta\chi'-m^2 \; \delta \chi - R\partial_R m^2 (\chi \delta\varphi + \varphi \delta \chi) &= -\Lambda \; \delta\chi,
	\end{split}
	\end{equation}
	where $\Lambda$ is the eigenvalue of the higher dimensional action, and by $(...)'$ we still mean $\rho$ derivatives. Comparing to \eqref{eq:metric_with_z} we learn that $\delta \varphi(\rho,z) = e^{i k z} \delta\varphi(\rho)$ and $\delta \chi(\rho,z) = e^{i k z} \delta\chi(\rho)$ are a solution for
	\begin{equation}
		\Lambda = \lambda + k^2.
	\end{equation}
	As $\lambda<0$, this higher dimensional mode is also negative for $k < \sqrt{-\lambda}$, and it becomes a solution for $k=\sqrt{-\lambda}$. The minimal possible wavelength is therefore $1/\sqrt{-\lambda}$.

	The existence of a $d$ dimensional negative mode $\lambda$ thus guarantees the existence of GL-type $9$ dimensional negative modes, as long as $1/\sqrt{-\lambda} \lesssim l_{ads}$. Because the solutions we found all have a length scale with $L \ll l_{ads}$, we expect this condition to hold at least far enough from $R_c$. At $R_c$ the negative mode seems to turn into a zero mode, and so close enough to $R_c$ no harmonic of $X_{9-d}$ could be excited. 

	It is interesting to go back and compare this result to the small AdS black hole saddle. As we mentioned above, the GL instability is believed to describe the dynamical decay of the small AdS black hole to $10$ dimensional black holes (see figure \ref{fig:phase_map}). We interpret the modes we just found as the avatar of the GL instability in the string star phase. What is than the analog of the $10$ dimensional black holes around $R_0 \sim R_H$? In other words, what does the string star thermodynamically decay \emph{to}? At these temperatures the $10$ dimensional black hole solution can't be trusted, just like the small AdS black hole.
	One might naively expect a $10$ dimensional (flat space) string star as the answer, but such saddles don't seem to exist. We mention again the results of \cite{Chen:2021emg}, that large $D=d+1$ string size black holes might exist also close to and above the Hagedorn temperature. Perhaps $D=10$ is large enough, and the solution decays to string-size $10$ dimensional black holes?

\section{Numerical results} \label{sec:numerical_res}
	\begin{figure}[t]
		\centering
		\includegraphics[width=.8\linewidth]{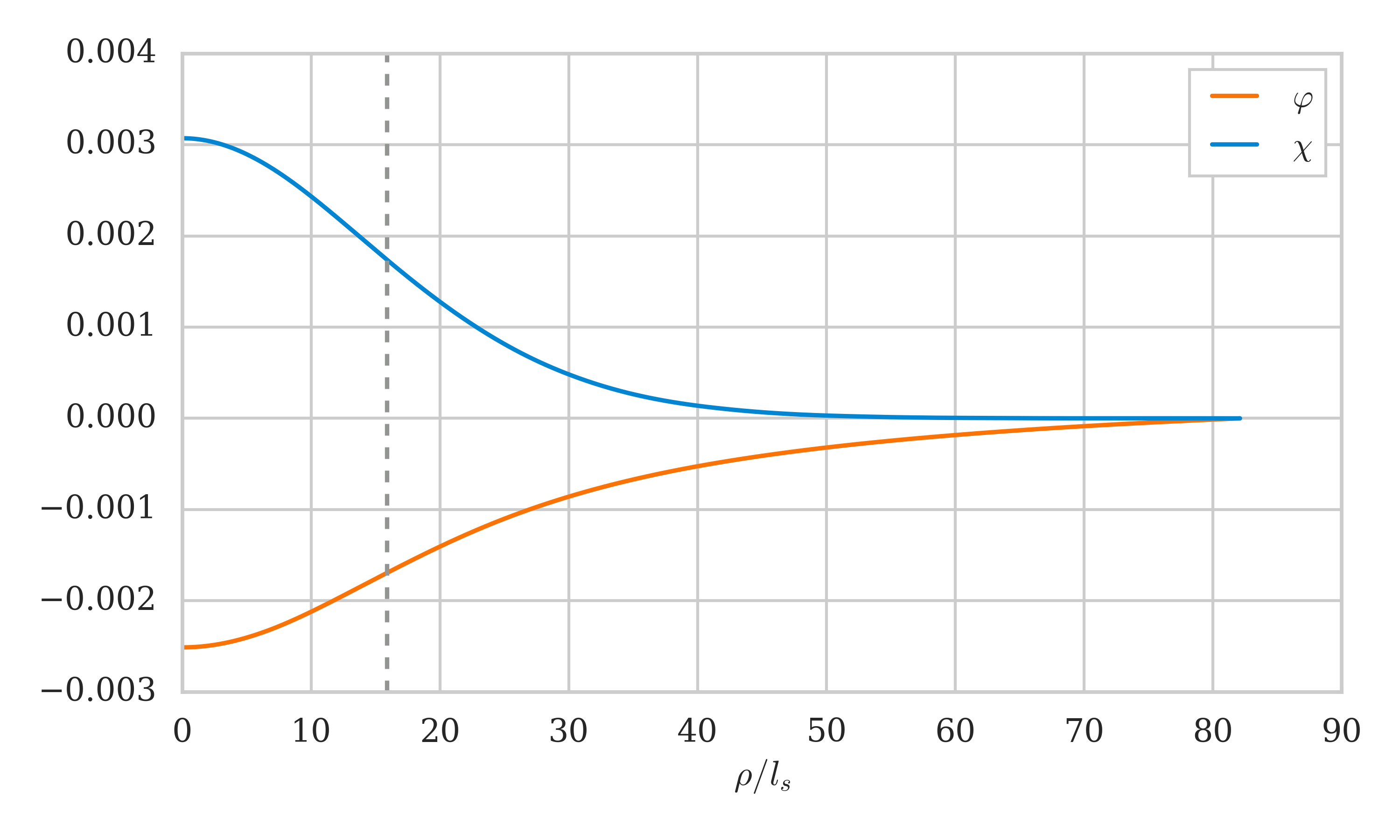}
		\caption{\label{fig:num_profile}
		An example of typical $\varphi(\rho)$, $\chi(\rho)$ profiles as a function of $\rho$ (in string units). The numerical calculation was done for the $d=3$ heterotic string with $l_{ads}/l_{s}=600$ at the (flat space) Hagedorn temperature $R_H=\left(1+1/\sqrt{2}\right) l_s$. 
		The gray vertical line is $\rho = L_H$, found in section \ref{sec:at_hagedorn}.}
	\end{figure}
	In order to study the string saddle beyond the analytical approximations, we used numerical simulations to find the behavior of the saddles as a function of $R_0$.
	We assumed a real and spherical solution for $\chi,\varphi$ and set to solve the coupled equations \eqref{eq:eom_rho}. Mathematically, the boundary conditions are $\varphi'(0)=\chi'(0)=0$, for smoothness at $\rho=0$, and $\varphi(\infty)=\chi(\infty)=0$. The simulation ran in string units $l_s=1$, which we will use below. In order to regularize the calculation, we arbitrarily chose $\rho_\text{min}=0.01$ and $\rho_\text{max}$ (which will be determined shortly), and defined the regularized boundary conditions
	\begin{equation}\label{eq:BC}
		\varphi'(\rho_\text{min})=\chi'(\rho_\text{min})=0,\qquad \varphi(\rho_\text{max}) = \chi(\rho_\text{max})=0.
	\end{equation}
	To find the solution of \eqref{eq:eom_rho} under the regularized boundary conditions, we used the shooting method: we trade the boundary condition at $\rho_\text{max}$ by a condition on the values of $\varphi_0 = \varphi(\rho_\text{min})$ and $\chi_0 = \chi(\rho_\text{min})$. For a given $\rho_\text{max}$, we optimize the values of $\varphi_0,\chi_0$ to solve \eqref{eq:BC}. We would like to take $\rho_\text{max}$ to be as large as possible. We searched for the largest $\rho_\text{max}$ for which we could still find $\varphi_0,\chi_0$ sufficiently well. In practice, $\rho_\text{max}$ is of the same order of magnitude as the solution's length scale $L$.

\begin{figure}[t]
		\centering
		\includegraphics[width=.8\linewidth]{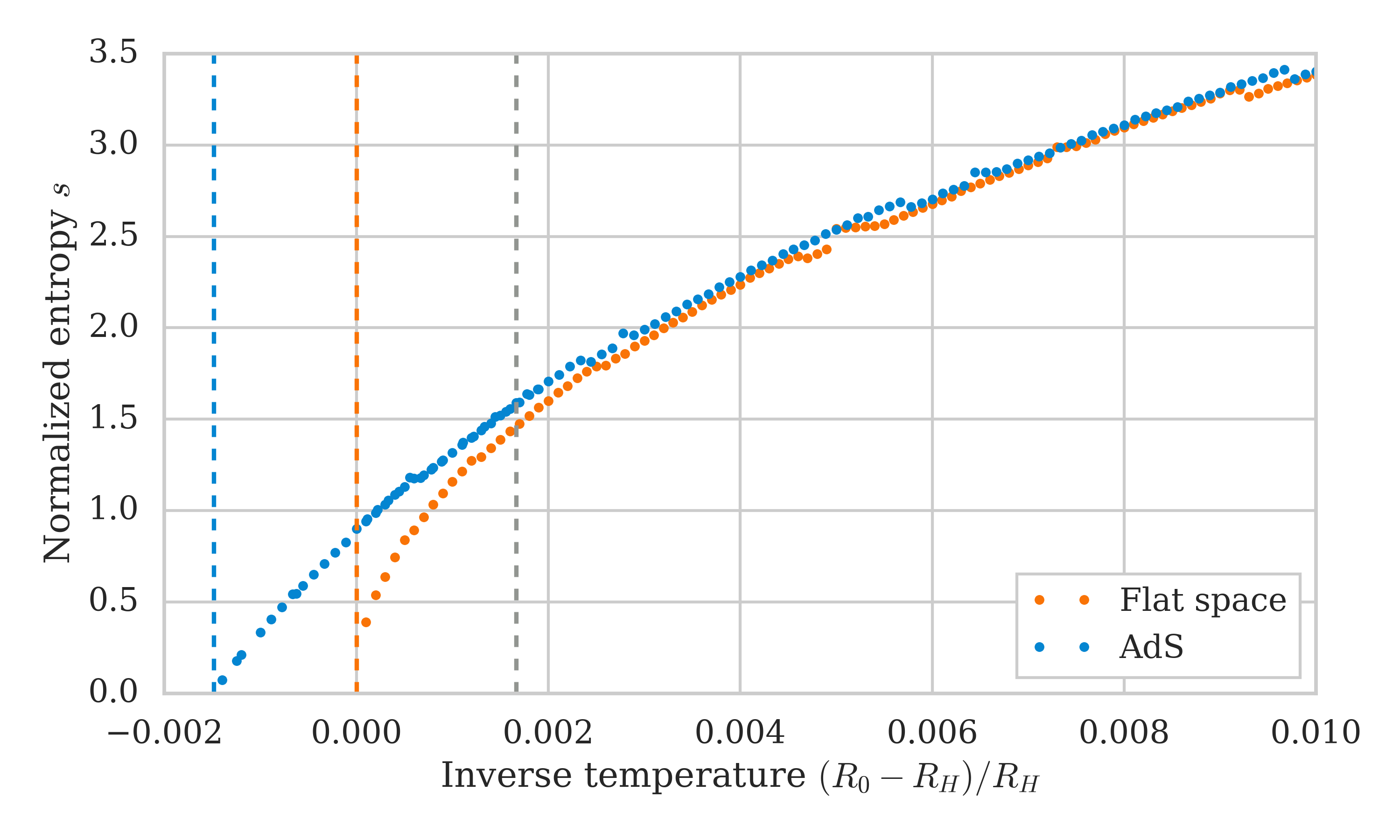}
		\caption{\label{fig:entropy_num}
		Numerical estimation of the string star normalized entropy $s$ as a function of $(R_0-R_H)/R_H$. The numerical calculation was done for the $d=3$ heterotic string. In blue is the AdS solution with $l_{ads}/l_{s}=600$, and in orange the flat space solution. 
		The gray vertical line is the temperature scale around which the AdS solutions significantly deviate from flat space. The orange vertical line is the flat space Hagedorn temperature $R_H$, and the blue vertical line is the AdS Hagedorn temperature $R_c$.}
	\end{figure}

\begin{figure}[t]
		\centering
		\includegraphics[width=.8\linewidth]{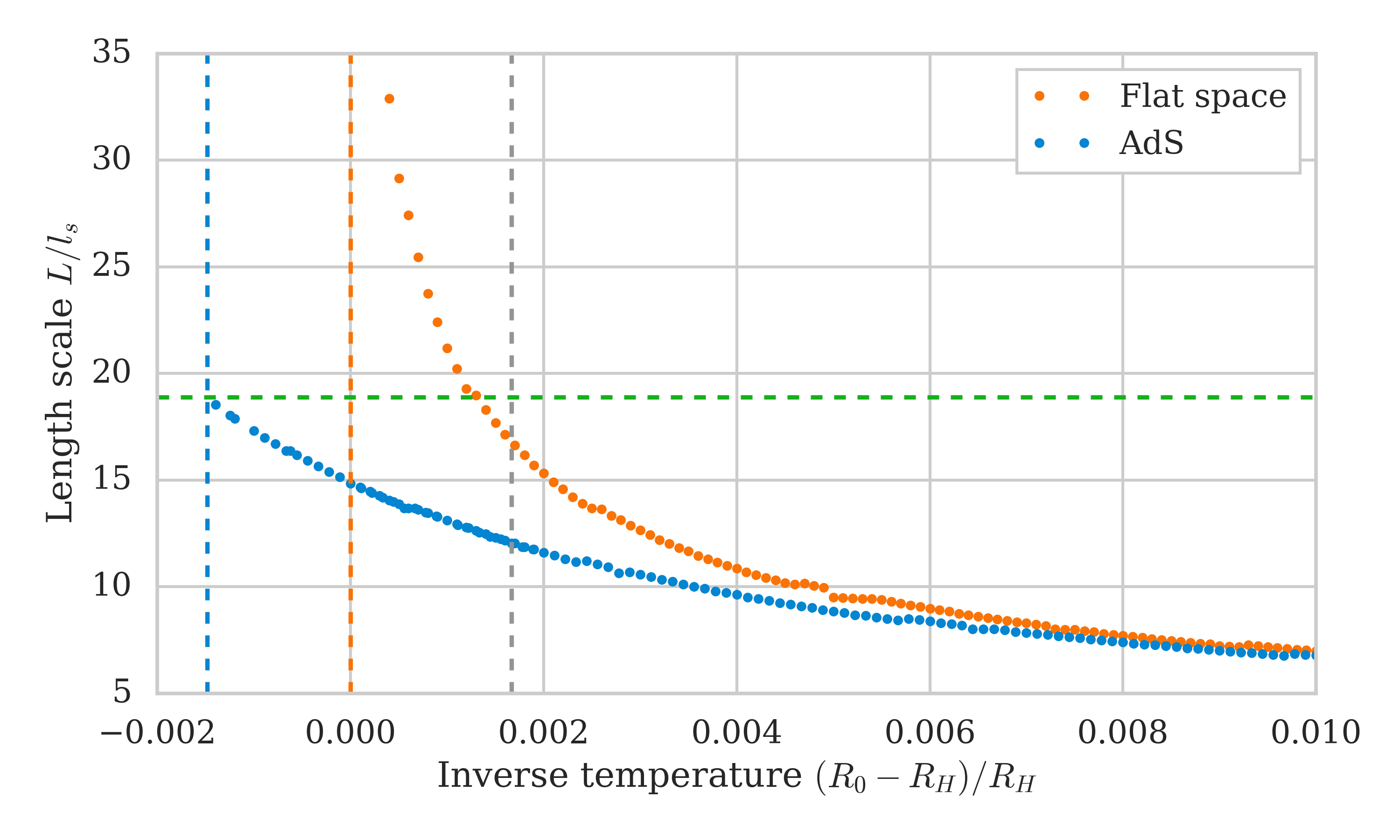}
		\caption{\label{fig:length_scale_num}
		Numerical estimation of the string star length scale $L$ (in string units) as a function of $(R_0-R_H)/R_H$. The length scale is defined as the coordinate $\rho$ at which $\chi(\rho)$ loses $1/\sqrt{e}$ of its amplitude at $\rho=0$. 
		The numerical calculation was done for the $d=3$ heterotic string. In blue is the AdS solution with $l_{ads}/l_{s}=600$, and in orange the flat space solution. 
		The gray vertical line is the temperature scale around which the AdS solutions significantly deviate from flat space. The orange vertical line is the flat space Hagedorn temperature $R_H$, and the blue vertical line is the AdS Hagedorn temperature $R_c$. The horizontal green line is the critical length scale $L_c$.}
	\end{figure}

\begin{figure}[t]
		\centering
		\includegraphics[width=\linewidth]{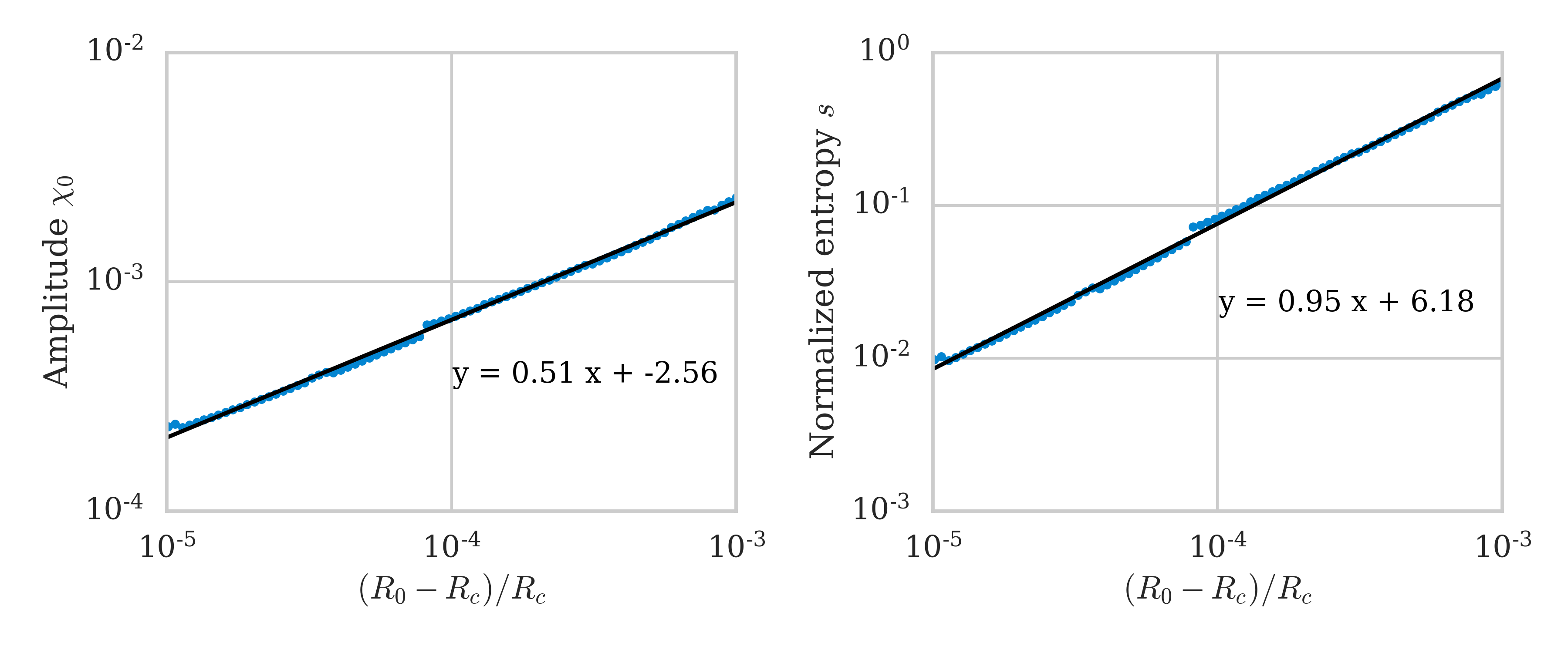}
		\caption[LoF entry]{\label{fig:critical_behavior_num}
		Loglog plots of the amplitude $\chi_0$ (left) and the normalized entropy (right), close to the critical temperature $R_c$ as a function of $(R_0-R_c)/R_c$. 
		The numerical results (in blue) are for the $d=3$ heterotic string and $l_{ads}/l_{s}=600$. In black, we draw a linear fit of the results. The same power-law exponent was found for other values of $l_{ads}$ as well.
		}
	\end{figure}

\begin{figure}[t]
		\centering
		\includegraphics[width=.8\linewidth]{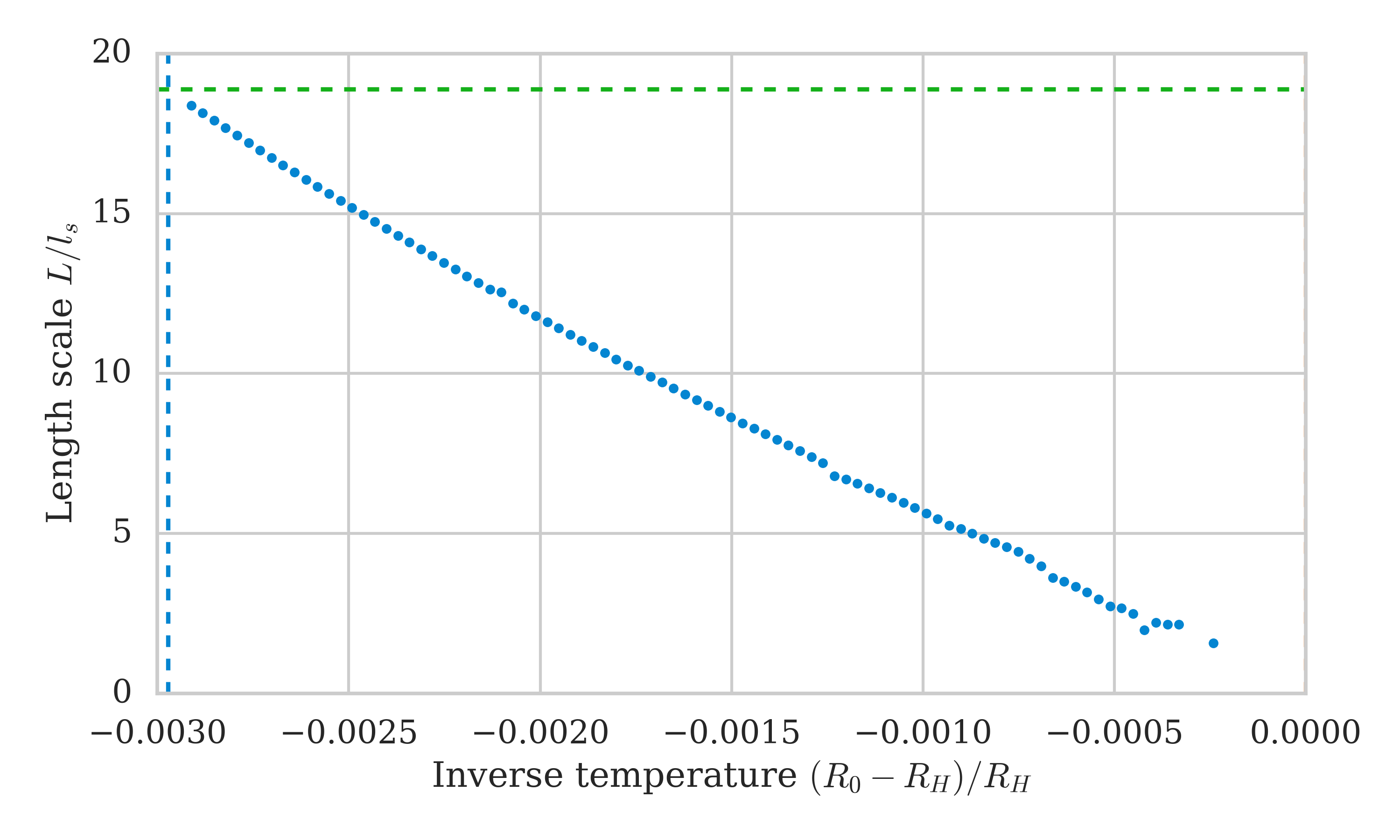}
		\caption[cap]{\label{fig:length_scale_num_6d}
		Numerical estimation of the AdS string star length scale $L$ (in string units) as a function of $(R_0-R_H)/R_H$.
		The numerical calculation was done for the $d=6$ heterotic string. The blue vertical line is the AdS Hagedorn temperature $R_c$. The horizontal green line is the critical length scale $L_c$.}
	\end{figure}

\begin{figure}[t]
		\centering
		\includegraphics[width=.8\linewidth]{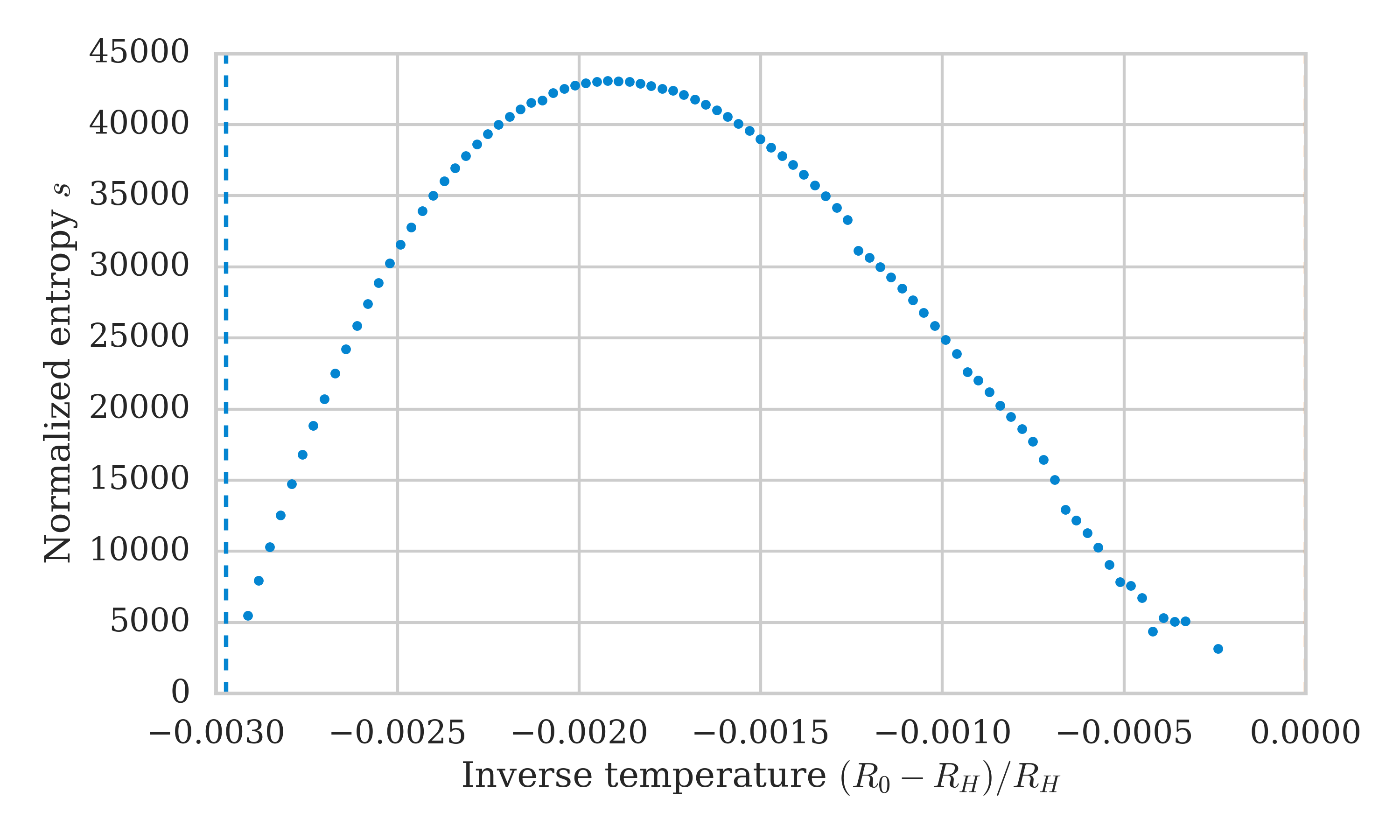}
		\caption[cap]{\label{fig:entropy_num_6d}
		Numerical estimation of the AdS string star normalized entropy $s$ as a function of $(R_0-R_H)/R_H$. The numerical calculation was done for the $d=6$ heterotic string. The blue vertical line is the AdS Hagedorn temperature $R_c$.}
	\end{figure}

	An example of a typical profile is given in figure \ref{fig:num_profile}.
	Figures \ref{fig:entropy_num} and \ref{fig:length_scale_num} plot the normalized entropy and the length scale of the numerical solutions as a function of $(R_0-R_H)/R_H$. Both figures include also the flat space simulations for comparison. Both are done for $d=3$ spatial dimensions and $l_{ads} = 600$ (in string units). The normalized entropy $s$ was calculated by numerical integration of \eqref{eq:entropy_def}. The length scale was defined as the coordinate $\rho = L$ at which $\chi(L)$ loses $1/\sqrt{e}$ of its original amplitude $\chi_0$. Figure \ref{fig:critical_behavior_num} shows the power-law behavior of the free energy and the amplitude $\chi_0$ close to $R_c$. 
	Figures \ref{fig:length_scale_num_6d} and \ref{fig:entropy_num_6d} also plot the the normalized entropy and the length scale, this time for the $d=6$ Heterotic string on AdS.

	Broadly speaking, the numerical results continuously interpolate between the different predictions of the previous sections without adding any noticeable features. For completeness, we list these results below:
	\begin{itemize}
		\item For $3\le d \le 5$ and in the temperature range $l_s/l_{ads} \ll (R_0-R_H)/R_H \ll 1$ the AdS and the flat space solutions coincide. As we show in section \ref{sec:small_sol}, the two start to disagree around $(R_0-R_H)/R_H \sim l_s/l_{ads}$.
		\item The length scale of the AdS solution $L$ grows with the temperature, and is bounded by $L_c$ (see section \ref{sec:gas_transition}). In figures \ref{fig:entropy_num}, \ref{fig:entropy_num_6d} the critical length scale $L_c$ is drawn as a green horizontal line.
		\item The solution joins with the trivial solution at $R_0=R_c$. In figures \ref{fig:entropy_num}, \ref{fig:length_scale_num}, \ref{fig:length_scale_num_6d}, \ref{fig:entropy_num_6d} $R_0=R_c$ is drawn with a blue vertical line.
		\item The power-law behavior of the amplitude $\chi_0$ and the normalized entropy $s$ around $R_0 = R_c$ are shown in figure \ref{fig:critical_behavior_num}. The linear fit (in black) gives an amplitude proportional to $(R_0-R_c)^{1/2}$, and an entropy proportional to $(R_0-R_c)^1$.
		Both are in agreement with the results of section \ref{sec:gas_transition}.
		\item Below the flat space Hagedorn temperature $R_0>R_H$ we found solutions only for $3\le d \le 5$ spatial dimensions, consistent with the flat space result \cite{Horowitz:1997jc,Chen:2021dsw}. Close to the AdS Hagedorn temperature $R_c$, however, we found solutions for $d\ge 6$ as well. Figures \ref{fig:length_scale_num_6d}, \ref{fig:entropy_num_6d} show the normalized entropy and the length scale of the string star solution for $d=6$. We can see in figure \ref{fig:length_scale_num_6d} that close enough to $R_c$ the solutions are reliable ($L \gg l_s$). The solution is unreliable close to $R_H$, as its size is around the string scale and its amplitude is $O(1)$. In figure \ref{fig:entropy_num_6d} we see that the normalized entropy there is expected to be $\sim O(1)$. Therefore around $R_0 \sim R_H$ the solution size is $L\sim l_s$ and its entropy is $S\sim l_s^{d-1}/G_N$, qualitatively agreeing with a string-sized (small) AdS black hole.
	\end{itemize}

\section{Holographic entanglement entropy of a string star} \label{sec:ee}

\begin{figure}[t] 
	\centering
	\includegraphics[width=.5\linewidth]{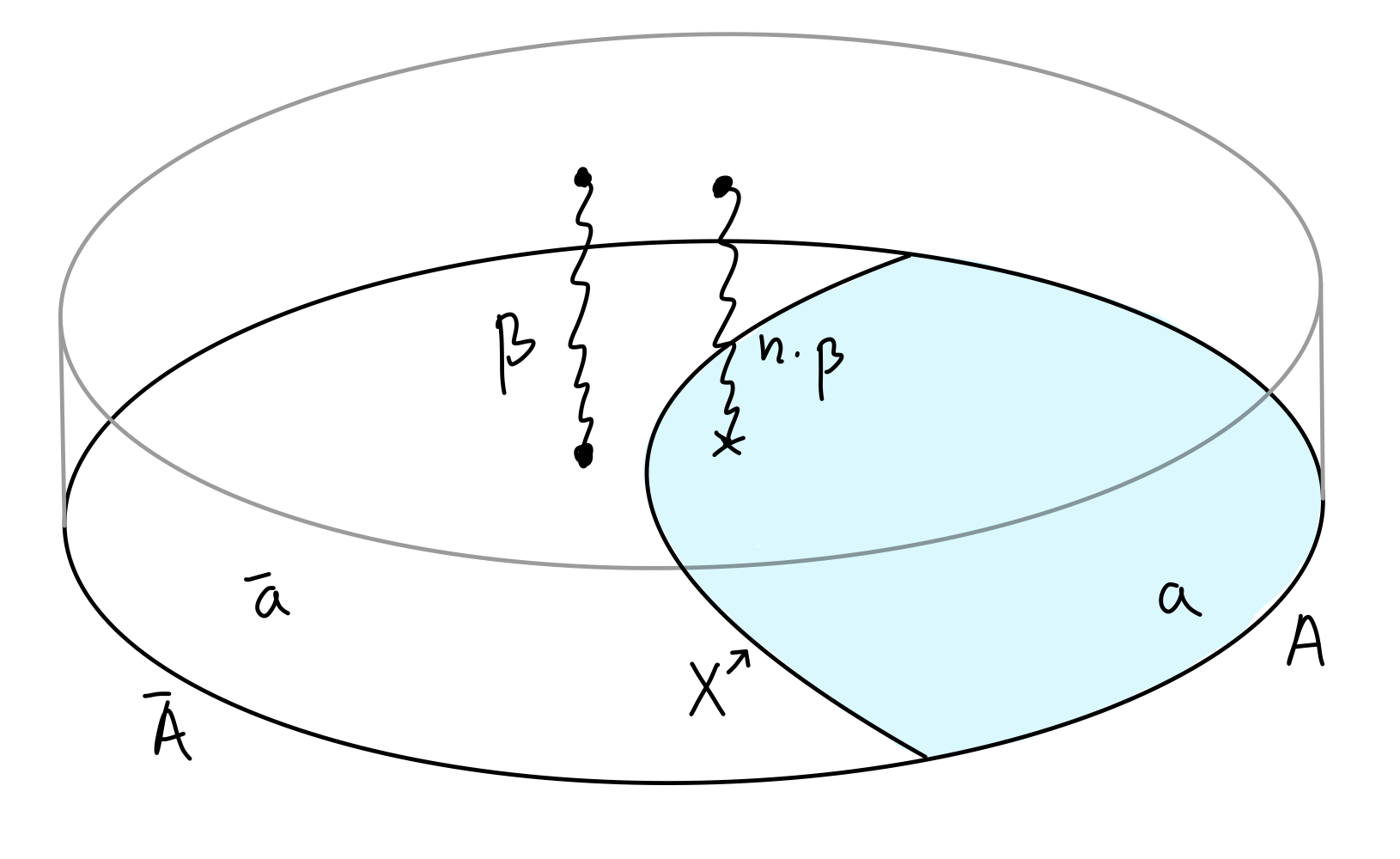}
	\caption{\label{fig:ee} The string theory $n$ replicas construction in thermal AdS, for near-horizon temperatures. In blue, the bulk replica surface $a$.
	The winding string in $a$ winds through all the $n$ replicas. As a result, the first winding mode has a mass $m^2(n \cdot R)$.
	For each replica, the first winding mode in the complement $\bar a$ has a mass $m^2(R)$.
	}
\end{figure}

In this work we only considered the thermodynamic properties of the string star, without discussing the corresponding stringy microstates. 
Yet it is still possible to describe its thermal mixed state using the Euclidean saddle we found. 
The density matrix is formally given by $\hat \rho = \frac{1}{Z} \exp(-\beta H_{SS})$, where $H_{SS}$ is the Hamiltonian projected to the string star states and $Z$ an appropriate normalization.
We stress that we are not offering here any way to identify this mixed state in the holographic CFT.
Notice that the explicit solutions for $\chi$ can be used as a bulk expression for the mixed state only after taking a trace and dimensionally reducing on the thermal circle. In this section we will use this description to find the holographic entanglement entropy of the string star thermal mixed state. 

We separate the CFT spatial $S^{d-1}$ into a region $A$ and its complement $\bar A$.
Following \cite{Lewkowycz:2013nqa} we employ the CFT replica trick in order to compute the entanglement entropy $S(A)$:
\begin{equation} \label{eq:entropies}
\begin{split}
	S(A) &= - \text{Tr}(\hat \rho_A \log \hat \rho_A) = \lim_{n\rightarrow 1} S_n(\hat \rho_A),\\
	S_n(\hat\rho_A) &= \frac{1}{1-n} \log \text{Tr} \hat\rho_A^n = \frac{1}{n-1}(I_n-n \cdot I_1).
\end{split}
\end{equation}
In the first line $\hat \rho_A = \text{Tr}_{\bar A} \hat\rho$ is the reduced density matrix. In the second line we defined the \myrenyi entropy $S_n$. On the RHS we define $I_n = -\log Z_n$, where $Z_n$ is the CFT n-replica path integral \cite{Rangamani:2016dms}. Using the holographic dictionary, we can compute $Z_n$ using a saddle point approximation. The calculation is entirely Euclidean, albeit over a complicated manifold. We assume that our definition of $\hat\rho$ instruct us to look (at least when $n\approx 1$) for an Euclidean string star solution in (almost) thermal AdS space.

Solving Einstein equations with the replicated boundary conditions of $Z_n$ gives a bulk geometry (close to $n=1$) of $n$ replicated $d+1$ dimensional thermal AdS spaces. The geometry is depicted in figure \ref{fig:ee}. We call the replicated $d$ dimensional submanifold $a$, and its complement $\bar a$. The boundary of $a$ in the bulk is called $X$, while the conformal boundary of $a$ is $A$. In this replicated Euclidean geometry we can take the dimensional reduction of the the thermal circle. In each of the $n$ replicas, the region $\bar a$ have a string winding mode with mass $m^2(R)$ \eqref{eq:m2_def}. Because of the replicated submanifold, the combined $a$ region can be understood as a spatial slice of thermal AdS with temperature $n \cdot \beta$. Alternatively, winding strings on $a$ extend through all the $n$ replicas and has a mass $m^2(n \cdot R)$ (see figure \ref{fig:ee}). For $n \approx 1$ and $\beta-\beta_H \ll l_s$ these modes are all light close enough to the origin. The resulting $d$ dimensional effective theory admits a bound state of the different winding modes (together with gravity). As we take the limit $n=1$ the classical solution merges with the solution we studied in this work.

We conclude that close enough to $n \approx 1$ the on-shell Euclidean action of the $n$-replicas string star solution is given by
\begin{equation} 
	I_n = n \cdot I(\bar a, \beta) + I(a, n \cdot \beta),
\end{equation}
where $I(a,\beta)$ is the on-shell integral \eqref{eq:action_d} with its domain restricted to $a$ and temperature $\beta$. This expression can be analytically continued in $n$. Substituting inside \eqref{eq:entropies} and following the arguments of section \ref{sec:thermo} we find the final result
\begin{equation} \label{eq:ss_entropy}
	S(A) = \frac{\text{Area}(X)}{4G_N} + \frac{1}{16\pi G_N} \int_a d^d x \sqrt{g^{(d)}} (2\pi R) \left( (R\partial_R m^2) + ((R\partial_R)^2m^2) \; \varphi \right) |\chi|^2 + O(G_N^0).
\end{equation}
Here $\chi$ and $\varphi$ are the original ($n=1$) solution of the string star.
The surface $X$ should extremize the entire functional, and thus slightly affected by the matter term \cite{Engelhardt:2014gca}.

There's a simple explanation for this formula. For a general matter (mixed or pure) state in AdS, the entanglement entropy is (at leading order in $\alpha'$) \cite{Faulkner:2013ana}
\begin{equation}\label{eq:RT}
	S(A) = \frac{\text{Area}(X)}{4G_N} + S^{bulk}(a),
\end{equation}
where $X$ is the appropriate extremal surface, and $S^{bulk}(a)$ is the bulk matter entanglement entropy in $a$. As $\chi(x)$ supposedly stands for a ($l_s$ sized) thermal stringy state around the point $x$, our state is a local mixed thermal state. The integrand of \eqref{eq:entropy_def} should thus be understood as a thermal entropy density. As a result the formula \eqref{eq:RT} immediately gives \eqref{eq:ss_entropy}. Note that this effective description requires $L \gg l_s$ (see also \cite{He:2014gva,Balasubramanian:2018axm,Chandrasekaran:2021tkb}), as we made sure in the previous sections. 

In light of the correspondence principle, we would like to compare the entanglement entropy of the string star \eqref{eq:ss_entropy} to that of a small AdS black black hole, given by \cite{Marolf:2018ldl,Ryu:2006bv}
\begin{equation} \label{eq:BH_RT}
	S_\text{BH} = \frac{\text{Area}(X_\text{BH})}{4G_N}+ O(G_N^0),
\end{equation}
where $X_\text{BH}$ is the extremal surface in the background of a small Euclidean AdS black hole. For the comparison we take the CFT region $A$ to be a polar-cap of the boundary $S^{d-1}$ covering angles from $0$ to $\theta_A$, parametrized by $\theta_A \in [0,\pi]$.
For small AdS black holes $\text{Area}(X_\text{BH}) \approx \text{Area}(X_\text{vac})+ \delta A$, $X_\text{vac}$ is the standard thermal AdS extremal surface \cite{Rangamani:2016dms}. $\delta A$ increase monotonically as a function of $\theta_A$ from zero at $\theta_A=0$ to the horizon area $\sim \beta^{d-1}$ at $\theta_A=\pi$. Due to the small size of the black hole, the majority of the increase happens around 
$|\theta_A -\frac{\pi}{2}| \sim \beta/l_{ads}$.
Considering the string star, the area term in \eqref{eq:ss_entropy} is $X\approx X_\text{vac}$ to leading order, with a correction at the order of the matter term. The matter term itself is also monotonic in $\theta_A$, from zero at $\theta_A=0$ to its full value \eqref{eq:entropies} at $\theta_A=\pi$. For low temperatures and $3\le d \le 5$ (see section \ref{sec:small_sol}) the maximal value is $\sim \frac{l_s^{d-1}}{G_N} \cdot \left((\beta-\beta_H)/\beta_H \right)^{\frac{4-d}{2}}$. The majority of the increase happens around 
$|\theta_A -\frac{\pi}{2}| \sim L/l_{ads}$, 
with the (low temperature) length scale $L / l_s\sim \left((\beta-\beta_H)/\beta_H \right)^{-\frac{1}{2}}$. 
We learn that the extrapolations of $S(A)$ of the small AdS black hole and the AdS string star to $\beta-\beta_H \sim l_s$ agree qualitatively, by the same reasoning as in section \ref{sec:small_sol}.

It is interesting to see that in terms of the CFT entanglement entropy, the area term appears both for the string star \eqref{eq:ss_entropy} and for the black hole \eqref{eq:BH_RT}. The string star entropy \eqref{eq:ss_entropy} includes another term that should be understood as both the ``matter contribution", in terms of \eqref{eq:RT}, but also as a correction of the ``tree level" area term. If the transition from the string star to the black hole is indeed continuous, this term continuously joins with the black hole area term.
Finding the generalization of \eqref{eq:RT} to finite $\alpha'$ (but at leading $1/G_N$ order) might help in understanding the relation between \eqref{eq:ss_entropy} and \eqref{eq:BH_RT}.

\section*{Acknowledgements}

I would like to thank Micha Berkooz, Nadav Brukner, Rohit Kalloor, Zohar Komargodski, David Kutasov, Ohad Mamroud, Adar Sharon, Tal Sheaffer, Masataka Watanabe, and Yoav Zigdon for useful discussions, and especially Ofer Aharony for many helpful discussions and guidance. I also thank Ofer Aharony, Yiming Chen, David Kutasov, Juan Maldacena and Adar Sharon for their comments on the manuscript.
This work was partly funded by an Israel Science Foundation center for excellence grant (grant number 2289/18), by grant no. 2018068 from the United States-Israel Binational Science Foundation (BSF), by the Minerva foundation with funding from the Federal German Ministry for Education and Research, by the German Research Foundation through a German-Israeli Project Cooperation (DIP) grant ``Holography and the Swampland", and by a research grant from Martin Eisenstein.

\appendix

\section{The effective field theory} \label{app:EFT}
Considering only the (string frame) metric $G$ and the dilaton $\Phi$ (which are universal), the $D=d+1$ dimensional low energy (Euclidean) action of string theory is \cite{Polchinski:1998rq} 
\begin{equation}
	I_{D} = \frac{1}{16\pi G_N} \int d^D x \sqrt{G} e^{-2\Phi}\left(-\mathcal{R} -4 \partial_\mu \Phi \partial^\mu \Phi \right).
\end{equation}
% In this normalization of $\Phi$, we can always rescale the metric by $G_{\mu \nu} = \exp\left(-\frac{4}{D-2}\Phi\right)\tilde G_{\mu \nu}$ which gives
% \begin{equation}
% 	I_{D} = \frac{1}{16\pi G_N} \int d^D x \sqrt{\tilde G} \left(-\tilde{\mathcal{R}} 
% 	+ \frac{4}{D-2} \partial_\mu \Phi \partial^\mu \Phi \right)
% \end{equation}
% In this form it is clear that we can always find solutions with $\Phi=\text{const}$ (which can be swollen inside the definition of $G_N$).

We would like to compactify the time direction and find the effective $d$ dimensional action. We will assume all the higher KK modes (including the massless vector) vanish and substitute a $U(1)$ invariant metric
\begin{equation}
	ds^2 = g_{ij}(x) dx^i dx^j + G_{tt}(x) dt^2.
\end{equation}
with $t\sim t+1$. The only non-zero components of the Christofell tensor are ${\Gamma^i}_{jk}$ and ${\Gamma^i}_{tt}=-\frac{1}{2}\partial^j G_{tt}$. As a result the Ricci scalar is
\begin{equation}
	\mathcal{R}_{D} = \mathcal{R}_{d} -\frac{1}{2} G^{tt} \nabla^2 G_{tt},
\end{equation}
with $\nabla^2$ the $d$ dimensional Laplacian. Substituting $G_{tt} = g_{tt} e^{2\varphi}$ and assuming the $D$ dimensional metric $g$ satisfies the Einstein equations (terms linear in $\varphi$ vanish), gives
\begin{equation}
	\mathcal{R}_{D} = \mathcal{R}_{D} \mid_{\varphi=0} - 2 \partial^i \varphi \partial_i \varphi.
\end{equation}
Substituting, the $d$ dimensional action is
\begin{equation}
	I_{d} = \frac{1}{16\pi G_N} \int d^d x \sqrt{ g}\sqrt{g_{tt}} e^{-2\Phi+\varphi}\left(-\mathcal{R}\mid_{\varphi=0} + 2 (\nabla \varphi)^2
	-4 (\nabla \Phi)^2 \right).
\end{equation}

When it is light enough, we can consistently add the first winding mode $\chi(x)$ to get
\begin{equation}
	I_{d} = \frac{1}{16\pi G_N} \int d^d x \sqrt{ g}\sqrt{g_{tt}} e^{-2\Phi+\varphi}\left(-{\mathcal{R}}\mid_{\varphi=0} + 2 (\nabla \varphi)^2
	-4 (\nabla \Phi)^2
	+|\nabla\chi|^2 +m^2(x) |\chi|^2
	\right).
\end{equation}
Here $m^2(R)$ is defined in \eqref{eq:m2_def}, with $R=\frac{1}{2\pi} \sqrt{g_{tt}}$.

To get the standard string-metric normalization we redefine $\phi_d = \Phi-\frac{\varphi}{2}$, which finally gives
\begin{equation}\label{eq:d_dim_action}
	I_{d} = \frac{1}{16\pi G_N} \int d^d x \sqrt{ g}\sqrt{g_{tt}} e^{-2\phi_d}\left(-{\mathcal{R}}\mid_{\varphi=0} + (\nabla \varphi)^2
	-4 (\nabla \phi_d)^2
	+|\nabla\chi|^2 +m^2(x) |\chi|^2
	\right).
\end{equation}

\section{The radial Green's function} \label{app:greens}
	\subsection{Flat space}
	The flat space radial Green's function $g(r,r')$ satisfies
	\begin{equation}
		\partial_r^2g(r,r')+\frac{d-1}{r}\partial_r g(r,r') = \delta(r-r')
	\end{equation}
	The general solution is (for $d>2$)
	\begin{equation}
		g(r,r') = \begin{cases}
		\frac{a_1(r')}{r^{d-2}}+a_2(r'), & r<r'\\
		\frac{b_1(r')}{r^{d-2}}+b_2(r'), & r>r'\\
		\end{cases}
		.
	\end{equation}
	Continuity of $g$ at $r=r'$ and discontinuity in $\partial_r g$ at $r=r'$ gives 
	\begin{equation}
	\begin{split}
		\frac{a_1(r)-b_1(r)}{r^{d-2}}+a_2(r) -b_2(r)&= 0,\\
		\frac{d-2}{r^{d-1}}(a_1(r)-b_1(r)) &= 1.
	\end{split}
	\end{equation}
	It defines $g(r,r')$ up to a homogeneous solution of $r$. The boundary conditions $g(r\rightarrow\infty,r')=0$ and $\partial_r g(r=0,r')=0$ fix the residual freedom, and give
	\begin{equation} \label{eq:flat_green}
		g(r,r') = (r')^{d-1}\cdot
		\begin{cases}
			\frac{(r')^{2-d}}{2-d}, & r<r'\\
			\frac{r^{2-d}}{2-d}, & r>r'\\
		\end{cases}.
	\end{equation}

	\subsection{AdS}
	We work with $l_{ads}=1$ and a radial coordinate $r$. The radial Green's function $k(r,r')$ satisfy
	\begin{equation}
	\begin{split}
		k''(r) + v(r) k'(r) &= \delta(r-r'),\\
		v(r) &= \tanh(r)+(d-1)\coth(r).
	\end{split}
	\end{equation}
	The general solution is
	\begin{equation}
		k(r,r') = 
		\begin{cases}
		a_1(r') + a_2(r') B\left(\text{sech}^2(r);\frac{d}{2},1-\frac{d}{2}\right), & r<r'\\
		b_1(r') + b_2(r') B\left(\text{sech}^2(r);\frac{d}{2},1-\frac{d}{2}\right), & r>r'
		\end{cases},
	\end{equation}
	with $B(x;a,b)$ the incomplete beta function.
	The boundary conditions are
	\begin{equation}
	 	\partial_r k(r= 0,r')=0, \qquad k(r\rightarrow \infty,r')=0,
	\end{equation}
	and continuity of $k$ at $r=r'$ and discontinuity in $\partial_r k$ at $r=r'$ give the solution
	\begin{equation}\label{eq:k_def}
		k(r,r') = -\frac{1}{2}\sinh^{d-1}(r')\cosh(r')\cdot
		\begin{cases}
			B\left(\text{sech}^2(r');\frac{d}{2},1-\frac{d}{2}\right),& r<r'\\
			B\left(\text{sech}^2(r);\frac{d}{2},1-\frac{d}{2}\right),& r>r'
		\end{cases}.
	\end{equation}

\bibliographystyle{JHEP}
\bibliography{ref}
\end{document}